\documentstyle[12pt,aaspp4,flushrt]{article}
\doublespace
\begin{document}

\title{Thermal X-Ray Pulses Resulting From Pulsar Glitches} 
\author{Anisia P.S.Tang and K.S. Cheng} 
\affil{Department of Physics, University of Hong Kong, Pokfulam Road, Hong Kong, China \\
e-mail address: hrspksc@hkucc.hku.hk}

\begin{abstract}
The non-spherically symmetric transport equations and exact thermal evolution model are used to calculate the transient thermal response to pulsars. The three possible ways of energy release originated from glitches, namely the `shell', `ring' and `spot' cases are compared. The X-ray light curves resulting from the thermal response to the glitches are calculated. Only the `spot' case and the `ring' case are considered because the
 `shell' case does not produce significant modulative X-rays. The magnetic field ($\vec B$) effect, the relativistic light bending effect and the rotational effect on the photons being emitted in a finite region are considered. Various sets of parameters result in different evolution patterns of light curves. We find that this modulated thermal X-ray radiation resulting from glitches may provide some useful constraints on glitch models.
\end{abstract}
\keywords{dense matter-stars: evolution-star: interiors-stars: neutron-stars: X-rays}

%%%%%%%%% Introduction %%%%%%%%%%%%%%%%%%%%%%
\section{Introduction}
Pulsar glitches are believed to be mainly either starquake-driven or superfluid-driven. The starquake mechanism was introduced by Ruderman (1969). He suggested that glitches could be caused by gravity. As a pulsar spins down due to dipole radiation, centrifugal force on the crust decreases and gravity pulls the crust towards a less oblate shape. Since pulsars are believed to be neutron stars with a solid crust, subsequent change in stellar shape induces stress in the crust until the maximum yield strength is reached. Sudden relaxation of this stress brings the stellar shape to equilibrium. The glitch energy is to be released in a small volume at the weak regions in the solid crust. This leads to the `spot' case as proposed by Van~Riper~\emph{et~al.}\ (1991). This localized heating process causes an uneven heating of the surface. As the pulsar rotates, the area of thermal photon emission facing the observer is changing. Together with the gravitational bending effect, the emission of thermal X-ray due to glitches will be at a particular phase in the light curve. Hence a periodic modulation of the  X-ray light curve will be observed. 

The superfluid-driven glitch mechanism was proposed by Anderson (1975). When the pulsar spins down, the vortices become unpinned from the lattice. They corotate with the local superfluid and scatter with the nearby lattices. There are two effects. One is superfluid angular momentum being transferred to the crust so that the crust spins up. The other is frictional heating being produced and causing local energy dissipation. Such a process occurs in a ring structure at the rotational equator (Alpar,~Anderson,~Pines~and~Shaham 1984). The detailed study of glitch energy depositing in a ring structure has been studied by other authors (eg.\ Bildsten~and~Epstein 1989; Epstein~and~Baym 1992; Link~and~Epstein 1996; Jones 1998).

In reality, the `spot-like' case and the `ring-like' case may not necessarily be mutually exclusive. After crust-breaking, which will cause a `spot' case, the whole crust will oscillate briefly about its new equilibrium configuration (Baym~and~Pines 1971). The vortices which are originally pinned to the crust will `feel' this oscillation and become unpinned from the crustal lattice. These depinned vortices start to scatter outward and transfer their differential rotational energy to the crust. Therefore, the `ring-like' glitch can follow. As a result, we believe that the energy release in a glitch is mainly composed of two components, namely the release of elastic energy in a starquake and the differential rotational energy released due to the depinning of vortices. On the other hand, several authors (Van~Riper~\emph{et~al.}\ 1991; Chong~and~Cheng 1994; Hirano~\emph{et~al.}\ 1997) assume the glitch energy to be released in a spherical shell at a certain density inside the pulsar, namely the `shell' case, though this does not seem to be realistic. There are other possible origins of the `spot' case of energy release (e.g.\ Srinivasan~\emph{et~al.}\ 1990; Ruderman 1991a; Link,~Franco~and~Epstein 1998) and the `ring' case of glitch energy release (e.g.\ Bildsten~and~Epstein 1989; Epstein~and~Baym 1992; Jones 1998). 

It has been suggested that the transient X-ray emission resulting from glitches will provide a good method to determine the equation of state for pulsars (neutron stars). However, the most active glitching pulsars (e.g.\ Vela, PSR 1706-16) are young neutron stars whose interior temperature is high ($\sim 10^8\ K$), this makes the overall luminosity variation difficult to detect (Van~Riper~\emph{et~al.}\ 1991; Chong~and~Cheng 1994). Cheng,~Li~and~Suen (1998) argue that if a good fraction of glitch energy is released in a small volume, then instead of heating up the entire stellar surface, even a small fraction of glitch energy can heat up a small area of the stellar surface drastically. Therefore, although after a glitch the total X-ray intensity varies very little, a very distinctive hot spot may be detected. They suggest that by observing the transient X-ray pulses, the equations of state of neutron stars can be determined. However, they have not considered such important effects as the relativistic light bending (Pechenick~\emph{et~al.}\ 1983) and the magnetic field effect (Page 1995) which can significantly affect the intensity and the pulse shape of the transient X-rays resulting from glitches.

In Section~\ref{section:input}, we summarize the input physics and the relativistic non-spherical symmetric thermal transport and energy balance equations that are used in determining the cooling process following a glitch. In Section~\ref{section:compare}, we apply the scheme mentioned in Section~\ref{section:input} to the three cases, namely the `spot', `ring' and `shell' cases, and compare the so-obtained temperature and luminosity profiles.  In Section~\ref{section:modulation}, we discuss the magnetic field effect, the relativistic light bending effect and the rotational effect. In Section~\ref{section:numeric}, we calculate the expected periodic modification to thermal X-ray pulses that are emitted due to `spot-like' and `ring-like' glitches resulting from the effects mentioned in Section~\ref{section:modulation}. In Section~\ref{section:conclusion}, we briefly summarize our results and discuss the detectability of the thermal X-ray profile being modified due to pulsar glitches.

\section{Physics inputs \label{section:input}}
No matter what the origins of the glitches are, glitches can be simulated by energy deposition in particular regions in the pulsar. The excess energy propagates in all directions. In this paper, we would like to know what fraction of glitch energy, which gives the transient emission of electromagnetic radiation from the stellar surface, can be observed. In calculating the thermal afterglow effect of glitches, the properties of a neutron star considered to be the major factors affecting the energy flow are the equations of state, the composition and the initial temperature profile.

\subsection{Neutron star structure}
The structure of a neutron star can be constructed by the general relativistic hydrostatic equilibrium equation (Tolman-Oppenheimer-Volkoff equation)
\begin{equation}
\frac{dP}{dr}=-\frac{G(m+4\pi r^2P/c^2)(\rho +P/c^2)}{r(r-2Gm/c^2)} \label{eq:str}
\end{equation}
where $P$ and $\rho $ are the pressure and mass density at radius  $r$ respectively, and
\begin{equation}
m=\int ^r_04\pi r^2\rho dr \label{eq:m}
\end{equation}
is the gravitational mass inside radius $r$. $G$ and $c$ are the gravitational constant and the speed of light in a vacuum respectively.

We will only calculate the temperature profiles between the radii with densities between $10^9\ g/cm^3$ and the nuclear density $\rho _N$. From $10^9\ g/cm^3$ to the stellar surface, we assume that the region between these two densities reaches equilibrium quickly, so that the temperatures at these two densities are related by using the formula proposed by Gudmundsson~\emph{et~al.}\ (1983). In this paper, since different equations of state give rise to different kinds of structure for a neutron star, the major equation of state (EOS) used is UT, which is based on the combined Hamiltonians consisting of UV14 and TNI models (Largaris and Pandharipande 1981; Friedman~and~Pandharipande 1981; Wiringa~and~Fiks 1988). It is a moderately stiff equation of state, when compared to the softer BPS model of Baym~\emph{et~al.}\ (1971), and the stiffer PPS model of Pandharipande~\emph{et~al.}\ (1976).

\subsection{Composition}
The ion, neutron and proton mass fraction, electron fraction, the mass number and proton number in the crust region denoted by $X_{ion}$, $X_n$, $X_p$, $Y_e$, $A$ and $Z$ respectively are obtained from Lattimer~\emph{et~al.}\ (1985). They are important for calculating the neutrino emissivity, heat capacity and thermal conductivity as presented in the following subsections.

\subsection{Neutrino emissivity}
The most important neutrino emission process for pulsars with age $\gtrsim 10^3\ years$ is electron bremsstrahlung in the crust (Flowers~and~Itoh 1976, 1979) as
\begin{equation}
Q_\nu ^{ions}=2.1\times 10^{20}\frac{Z^2}{A}(\frac{\rho }{\rho _N})T^6_9\ erg\,cm^{-3}\,s^{-1}
\end{equation}
where $T_9$ is the temperature in units of $10^9\ K$ and $\rho _N=2.8\times 10^{14}\ g/cm^3$ is the nuclear density. Other neutrino emissivities (cf.\ Chong~and~Cheng 1994 for a brief review) are also included.

\subsection{Heat capacity ($C_v$)}
In the crust, there are extremely relativistic degenerate electrons, some non-relativistic neutrons and ions, but there are no free protons. The scheme for calculating heat capacities follows Chong~and~Cheng (1994) and Cheng,~Li~and~Suen (1998). 

\subsection{Thermal conductivity ($K$)}
Thermal conductivity is fitted according to the work of Itoh~\emph{et~al.}\ (1984) in solid phase and that of Itoh~\emph{et~al.}\ (1983) in liquid phase. In regions of $\rho >1.311\times 10^{14}\ g/cm^3$, linear extrapolation is used.

\subsection{Initial temperature}
Before a glitch occurs, the temperature of the pulsar should be in equilibrium. For young and middle-aged pulsars, in which glitches can occur, the core temperatures are about $\sim 10^7-10^8\ K$. Since different EOS give rise to different kinds of structure, the equilibrium temperature profile of a pulsar depends on the EOS. 

\subsection{Non-spherically symmetric general relativistic transport and energy balance equations}
According to Cheng~\emph{et~al.}\ (1998), the general relativistic thermal transport equation and the energy balance equation are given by
\begin{equation}
\frac{e^{-\Phi }e^{-\Lambda }}{r^2}\,\frac{\partial}{\partial r}\,(r^2F_re^{2\Phi})
+\frac{e^{\Phi}}{r \sin \theta} \frac{\partial}{\partial \theta}\,(\sin
\theta\,F_\theta ) 
+\frac{e^{\Phi}}{r \sin \theta} \frac{\partial}{\partial
\phi}\,F_\phi =-(C_v\,\frac{dT}{dt}+e^{\Phi} Q_\nu )
\end{equation}
and
\begin{eqnarray}
& \frac{\partial}{\partial r}\,(Te^{\Phi})\, & = -\frac{e^{\Phi} e^{\Lambda}}{K}\,F_r \\
& \frac{1}{r}\,\frac{\partial}{\partial \theta}\,(Te^{\Phi})\, & = -\frac{e^{\Phi}}{K}\,F_\theta \\
& \frac{1}{r \sin \theta}\,\frac{\partial}{\partial \phi}\,(Te^{\Phi})\, & = -\frac{e^{\Phi}}{K}\,F_\phi
\end{eqnarray}
where $T$ is the temperature, $e^\Phi $ and $e^\Lambda $ are the redshift factor and length correction factor respectively, and $F_r$, $F_\theta $ and $F_\phi $ are the heat fluxes along $r$, $\theta $ and $\phi $ directions respectively.

\subsection{Heat input}
As we have mentioned in the introduction, there are likely two types of energy release during the glitches, namely, the elastic energy of the crust (Baym~and~Pines 1971) and the differential rotation energy between the crustal superfluid and the solid crust. In the former case, the energy is released in a localized volume, which is referred to as the `spot' case. In the latter case, the energy is released in the equatorial plane and hence is called the `ring' case. The amount of energy released in these two cases is estimated as follows:

\subsubsection{`Spot' case}
The energy released in a starquake comes from the relief of strain energy and is estimated to be (Baym~and~Pines 1971; Ruderman 1991b; Cheng~\emph{et~al.} 1992)
\begin{equation}
\triangle E_{strain}\sim \mu V_{crust}\theta _{max}^2
\end{equation}
where $\mu $ [$\sim 10^{29}(\rho /10^{13}\ g/cm^3)^{-4/3}\ dynes\ cm^{-3}$ where $\rho $ is the mass density] is the mean shear modulus, $V_{crust}$ is the volume of the crust from which energy is released, and $\theta _{max}$ [$\sim 10^{-1}-10^{-2}$ for a pure coulombic lattice and $10^{-3}-10^{-4}$ for an impurity dominated lattice (Smolukowshi 1970)] is the maximum strain angle that the crust can withstand without cracking. If the glitch occurs at $\rho \sim 10^{13}\ g/cm^3$, the strain energy released is estimated as $\triangle E_{strain}\sim 10^{40}(\theta _{max}/10^{-2})^2\ erg$. Hence, the estimated $\triangle E_{strain}\lesssim 10^{40}\ erg$.

It has been argued that the magnetic pressure of magnetars, which are neutron stars with extremely strong magnetic fields ($\sim 10^{15}\ G$), is strong enough to break the crust during the evolution of magnetic field and may encourage glitching with an amplitude $\frac{\triangle \Omega }{\Omega }\sim 10^{-5}$ (Thompson~and~Duncan 1996). In fact, Heyl~and~Hernquist (1999) have found evidence for glitches occuring in possible magnetar candidates, IE1048.1-5937 and IE2259+586, with amplitude $\frac{\triangle \Omega }{\Omega }\sim 10^{-5}$. If a large fraction of magnetic energy is dissipated inside the crust, the heat dissipation resulting from the glitch can be estimated as $I_\star \Omega ^2(\frac{\triangle \Omega }{\Omega })\sim 10^{40}\ erg$ using the typical parameters of magnetars ($\Omega \sim 1\ rad/s$, $\frac{\triangle \Omega }{\Omega }\sim 10^{-5}$ and $I_\star \sim 10^{45}\ g\ cm^2$ where $I_\star $ is the total moment of inertia of the star).

\subsubsection{`Ring' case}
In a superfluid-driven glitch originated from a sudden transfer of angular momentum from the inner crust superfluids to the crust, the angular momentum loss for the crustal superfluid is $I_{cr}\delta \Omega _s$ where $I_{cr}$ is the moment of inertia for the crustal superfluid and $\delta \Omega _s$ is the angular velocity change of crustal superfluid before and after the glitch. The angular momentum loss for the charged component, including the stellar core, which is strongly coupled to the solid crust via electron-magnetized vortex scattering (Alpar,~Langers~and~Sauls 1984), and the solid crust, is $I_{ch}\triangle \Omega $, where $\triangle \Omega $ is the observed angular momentum jump of the glitch and $I_{ch}$ is the moment of inertia for the charged component. By angular momentum conservation $\triangle J=I_{ch}\triangle \Omega =I_{cr}\delta \Omega _s$ and the energy dissipated due to the loss of differential rotation energy between the crustal superfluid and the charged component is
\begin{equation}
\triangle E=\frac{1}{2}[I_{ch}\Omega ^2+I_{cr}\Omega _s^2]-\frac{1}{2}[I_{ch}(\Omega +\triangle \Omega )^2+I_{cr}(\Omega _s-\delta \Omega _s)^2]=\triangle J(\Omega _s-\Omega )\equiv \triangle J\Omega _{lag}.
\end{equation}
where $\Omega _{lag}$ is the angular speed difference between the crust and the superfluid. In the nuclear pinning region, $\Omega _{lag}$ is $\sim 1-100\ rad/s$ (Alpar,~Cheng~and~Pines 1989). In the interstitial pinning region, it is $\lesssim 0.1\ rad/s$ (Link~and~Epstein 1991). According to the postglitch relaxation (Epstein,~Van~Riper~and~Link 1992; Alpar,~Chau,~Cheng~and~Pines 1993), $I_{ch}\gg I_{cr}$, $\triangle J=I_{ch}\triangle \Omega =(I_\star -I_{cr})\triangle \Omega \simeq I_\star \triangle \Omega $. For a typical neutron star with a giant glitch $\triangle \Omega \sim 10^{-4}\ rads^{-1}$, the moment of inertia of the star is $I_\star \sim 10^{45}\ g\ cm^2$ and the estimated energy release in nuclear pinning regions is $\sim 10^{41}-10^{43}\ erg$, while that in the interstitial pinning regions is $\sim 10^{40}\ erg$.

In general, the starquake glitches (the `spot-like' cases) can occur anywhere within the crust where a coulomb lattice exists, whereas superfluid-driven glitches (the `ring' cases) can only occur in the inner crust where neutron superfluid and coulomb lattices coexist. Therefore, they must occur in the inner crust at densities between $10^{12}$ and $2\times 10^{14}\ g/cm^3$. For simplicity, the `ring' case is assumed to occur at $\rho _{glitch} \sim 10^{13}-10^{14}\ g/cm^3$. Other cases can occur at $\rho _{glitch}\sim 10^{12}-10^{14}\ g/cm^3$. The amount of energy liberated, $\triangle E$, is thus estimated to be between $\sim 10^{40}\ erg$ and $10^{43}\ erg$. 

\section{Comparison of `shell', `ring' and `spot' cases \label{section:compare}}
A finite difference method is used to solve the equations of the previous section. For the `shell' case, spherical symmetry can be confidently assumed. As for the `spot' case, the cell at which energy release takes place can be treated as a `pole'. The direction joining the centre of the pulsar and the `pole' is defined to be the $\vec z$-direction. Since the rotation of the pulsar is slow, the direction of rotational axis can be neglected and azimuthal symmetry can be assumed. For the `ring' case, the angular velocity direction is taken to be the $\vec z$-direction, so azimuthal symmetry can be assumed again. Therefore, a 2-dimensional grid with $N_r\times N_\theta $ cells is used, where $N_r$ and $N_\theta $ are, for computational convenience, taken to be 100 and 90 respectively.

For the `spot' case, the release of energy is in one cell of the grid. For the `shell' case, the release of energy is in a layer of the grid. Whereas for the `ring' case, the release of energy is around the equator with a height of $4^o$, i.e. $\sim 0.69\ km$ for a UT-star, which is roughly the thickness of the crust.

The main parameter used for comparison of these various cases, is the total surface luminosity as observed by an observer at infinity, i.e. $L_s^\infty $. As $L\propto T^4$, the increase in temperature is reflected in the luminosity profile.

The temperature profile for a `ring' case glitch is shown in Fig.~\ref{fig:UTC7E42R13ringTr}. The distribution of surface temperature with respect to the angle from the rotational axis is shown in Fig.~\ref{fig:UTC7E42R13ringTa}. The distribution of surface temperature with respect to the angle from the spot for the `spot' case is shown in Fig.~\ref{fig:UTC7E42R13spotTa}. The evolution for the `ring' case is faster than the `spot' case as presented in Cheng,~Li~and~Suen (1998), but it is slower than the `shell' case as presented in Chong~and~Cheng (1994).

Since the heat propagation time is proportional to the heat capacity, which is approximately proportional to the temperature, i.e. $C_v\propto T$, and inversely proportional to the thermal conductivity, which is approximately inversely proportional to the temperature, i.e. $K\propto \frac{1}{T}$, the time that the glitch energy requires to reach the stellar surface is roughly proportional to the temperature squared, i.e. $t\propto \frac{C_v}{K}\propto T^2$ (Hirano~\emph{et~al.}\ 1997). It can be seen that the `shell' case gives the fastest response to the glitch because the same amount of energy is deposited in a larger volume, the localized heating effect is averaged out. Therefore, the temperature in each particular area is not so high and the heat energy can reach the surface faster than in the other two cases. 

The duration of the pulse on the surface for the `ring' case is also between that for the `shell' case and the `spot' case. This is similar to Chong~and~Cheng (1994)'s `shell' case, with a larger amount of energy deposited. The pulse will last a longer time because more heat is deposited in a particular volume and this will increase the diffusion time required to the next cell. Therefore, the pulse due to a `spot-like' glitch will last longer, while the second shortest one is the `ring' case and lastly the `shell' case.

On the other hand, the peak luminosity for the `spot' case is the highest because the localized heating effect is transferred to the surface. Also, the increase in luminosity lasts the longest time for the `spot' case because the heat diffusion speed is the lowest (c.f.\ Fig.~\ref{fig:UTC7E42R13} and Fig.~\ref{fig:UTC8E42R13}). The increase in luminosity is smaller for a cooler core than for a hotter core. The heat content of a pulsar is significantly enhanced by a glitch if the energy liberated is large compared to the original heat content (Chong~and~Cheng 1994). This can be understood by the fact that luminosity is proportional to $T^4$. Hence the temperature difference is magnified. However, the fractional increase $\frac{L_s(max)-L_s(0)}{L_s(0)}$ of a cooler core is $\sim 1.3$ and only $\sim 0.014$ for a hotter core. Therefore, it is much easier to observe the transient X-ray pulse from a cooler star than from a hotter star.

In Fig.~\ref{fig:C7E42R13ringEOS}, we can see that a softer EOS like BPS gives rise to a faster response and a higher peak. This is because a softer EOS gives a thinner crust. Heat diffuses to the surface faster because the distance to the surface is smaller. The deposition of energy to the core will also be smaller and a larger amount of the released energy is transported to the surface.

\section{Thermal X-ray modification \label{section:modulation}}
Andersen~and~\"Ogelman (1997) have already proposed that thermal afterglow caused by transient energy releases in a neutron star can alter its X-ray pulse shape by heating a portion of the crust so that more thermal X-rays are emitted at a particular phase. By  comparing such kinds of changes in X-ray pulses with the model results, constraints on  glitch models can hopefully be made. In the previous section, the temperature profiles on the surface of the pulsar for various models are obtained. However, in order to calculate thermal X-ray light curves resulting from glitches the gravitational lensing effect and the magnetic field effect must be considered.

\subsection{The effect of magnetic field}
It is well-known that the thermal conductivity in a magnetized neutron star depends on the angle between the heat current and the magnetic field because both the heat conduction and the opacity coefficient depend on the magnetic field direction (Hernquist 1984; Yakovlev~and~Urpin 1980; Yokovlev 1982; Tsuruta 1986; Schaaf 1987,1988, 1990). The anisotropic conductivity in the crust creates a distribution of the temperature over the stellar surface. Fortunately, it has been shown that an approximate solution to the thermal diffusion equation that accounts for this magnetic field effect can be reduced to a one-dimensional problem and the surface temperature distribution is given by (e.g.\ Page 1995).
\begin{equation}
T^4_{s,ani}(\vec B,T_b,\Theta _B)=T^4_{s,iso}(\vec B=0,T_b,\Theta _B=0)(\cos ^2\Theta _B+\chi _0^4(\vec B,T_b)\sin ^2\Theta _B)
\end{equation}
with $\chi _0^4=K_\bot /K_\|  $, the ratio of the thermal conductivities perpendicular and parallel to the magnetic field, which according to Greenstein~and~Hartke (1983),  is assumed to be constant within the envelope. $\chi_0$ is further established as $\frac{T_{s,ani}(\Theta _B=90^o)}{T_{s,ani}(\Theta _B=0^o)}$ in Page (1995). $\Theta _B$ is the angle between the local field and the radial direction. Here, only the dipolar magnetic field case is considered. $T_b$ is the temperature at a density of $10^{10}\ g/cm^3$, $T_{s,ani}$ is the surface temperature of the anisotropic case and $T_{s,iso}$ is the surface temperature of the isotropic (no magnetic field) case.

\subsection{The effect of relativistic light bending}
 Harding~and~Muslimov (1998) modeled the soft thermal X-ray profile by noticing that the magnetic polar caps of a cooling neutron star are slightly hotter than the rest of the stellar surface because of the strong magnetic field effect on heat transport in the surface layers. A similar approach is used to analyze the `spot' case in this paper, but the origin of the excessive heating is different. In fact, in their case, a steady X-ray pulse is expected but in our case, a transient X-ray pulse occurs. In this paper, the effect of the magnetic field is ignored.

The following scheme of calculating the X-ray light curves resulting from the gravitational bending and the stellar rotation is adapted from Pechenick~\emph{et~al.}\ (1983). Various model temperature profiles of the `spot' case are incorporated to determine the energy flux emitted from different positions of the pulsar. Since a neutron star has a large mass and a small radius, it is necessary to consider the gravitational deflection of the emitted photons (c.f.\ Fig.~\ref{fig:delta}). When the photons are emitted at an angle $\delta $ from the observer's direction, it will seem to  the observer that they are emitted at an angle $\theta '$. The relationship between $\delta $ and $\theta '$ for different EOS are plotted in Fig.~\ref{fig:angle}. The `flat' curve, which ignores the gravitational bending effect, is used for comparison. A softer EOS gives rise to a stronger lensing effect. This relation is generated from the equations given by Pechenick~\emph{et~al.}\ (1983): 
\begin{equation}
\xi +\beta =\theta '
\end{equation}
\begin{equation}
\beta =\int ^{\frac{GM}{Rc^2}}_0[(\frac{D\xi c^2}{GM})^{-2}-(1-2u)u^2]^{-1/2}\,du
\end{equation}
\begin{equation}
\delta =\sin ^{-1}(\frac{D\sin \xi}{R}\sqrt{\frac{1-\frac{2GM}{Rc^2}}{1-\frac{2GM}{Dc^2}}})
\end{equation}
where $D$ is the distance to the observer and the angles are defined as in Fig.~\ref{fig:delta}.

To avoid the complex procedures in solving the above three equations, some approximations can be made. Since $\frac{2GM}{Dc^2}$ is extremely small, and $\xi $ is also very small, $\delta $ can be treated as $\frac{D\xi }{R}(1-\frac{GM}{Rc^2})$ and $\beta \approx  \int^{\frac{GM}{Rc^2}}_0\frac{D\xi c^2}{GM}(1+\frac{u^2}{2}(\frac{D\xi }{GM/c^2})^2)\,du$ which is approximately $\frac{D\xi }{R}+\frac{(\frac{GM}{Rc^2})^3}{6}(\frac{D\xi }{GM/c^2})^3=\frac{D\xi }{R}+\frac{1}{6}(\frac{D\xi }{R})^3$. As a result, $\theta '\approx  \beta \approx \frac{D}{R}\xi +\frac{1}{6}(\frac{D\xi }{R})^3$. Hence, $\theta '-\delta \approx \frac{GM}{Rc^2}\delta $. Actually, this is a good approximation for $\delta <60^o$.

Since the rotation speed of a pulsar is much less than the speed of light, it can be treated as a slowly rotating rigid body, in which case
\begin{equation}
\cos (\theta _0)=\sin (\gamma )\sin (\gamma _o)\cos (\Omega t)+\cos (\gamma )\cos (\gamma _o)\label{eq:ask}
\end{equation}
where $\theta _0$ is the angle between the centre of `spot' region and the observer, $\gamma $ and $\gamma _o$ are the position angles from the z-axis to the centre of `spot' region and the observer respectively as described in Fig.~\ref{fig:theta0}.
$\Omega t$ is the azimuthal angle at time $t$ and $\Omega $ is the stellar angular velocity. So far, we have considered the emission region as a point, however, in reality, the emission region is of a finite area. If the photon emission occurs at a surface area defined in a cone with an angle $\alpha $ at the stellar surface, the result is then generated from the following equations according to Pechenick~\emph{et~al.}\ (1983):
\begin{equation}
h[\theta ;\alpha ,\theta _0]=2\cos ^{-1}(\frac{\cos \alpha -\cos \theta _0\cos \theta }{\sin \theta _0\sin \theta })
\end{equation}
for $\theta $ between $\theta _0\pm \alpha $.
\begin{equation}
\theta (x)=\int^{\frac{GM}{Rc^2}}_0[x^{-2}-(1-2u)u^2]^{-1/2}\,du.
\end{equation}
$\theta _0$ is defined in the same way as in Eq.~\ref{eq:ask}. The relative brightness is
\begin{equation}
A(\theta _0;f,M/R,\alpha )=(1-\frac{2GM}{Rc^2})^2(\frac{GM}{Rc^2})^2\int^{x_{max}}_0f(\delta (x))h[x;\alpha ,\theta _0]x\,dx
\end{equation}
where $f(\delta )=1$ for isotropic emission, $f(\delta )=\cos \delta $ for enhanced emission and $f(\delta )=\sin \delta $ for suppressed emission, $\delta =\sin ^{-1}(x/x_{max})$. As $\xi $ is small, $\sin \xi $ is approximated to $\xi $, so $x=D\xi c^2/GM$ and
\begin{equation}
x_{max}=(Rc^2/GM)(1-2GM/Rc^2)^{-1/2}
\end{equation}
where the angles are as defined in Fig.~\ref{fig:theta0}.

The relationship between $A$ and $\Omega t$ is plotted in Fig.~\ref{fig:brightEOS} for an isotropic emission. A softer EOS gives a stronger lensing effect so that when the `spot' is nearly at the back ($\sim 170^o$), it can still be observed even though the relative brightness is low. For a UT-star, the bending effect can extend the observed angle to nearly $30^o$. With a larger cone of emission, the relative brightness is larger when the `spot' is facing the observer, but it does not have any effect on the maximum angle of deflection.

According to Pechenick~\emph{et~al.}(1983), the total energy flux observed is:
\begin{equation}
F_X=\sum I_0(\frac{R}{D})^2A(\theta _0;f,M/R,\alpha )
\end{equation}
where $I_0$ is the energy flux at the surface for different angles from the spin axis and includes the factors $\cos \delta $ and $\sum $, which denotes the summation of the contributions from each cell.

Including the gravitational bending, together with the slowly-rotating and the finite area effects,  the evolution of total brightness for a UT-star with $T_c=10^8\ K$, $\triangle E=10^{40}\ erg$ and $\rho _{glitch}=10^{11}\ g/cm^3$ is plotted in Fig.~\ref{fig:UTC8E40R11chp5f}. Other sets of parameters give similar shape light curves, except that the number of days to reach the maximum will change, as will the fractional increase in energy flux.

\section{Numerical results \label{section:numeric}}
Consider the fractions $f_F=\frac{F_{X_{max}}-F_{X_{min}}}{F_{X_{min}}}$ and $f_T=\frac{T_{s_{max}}-T_{s_{min}}}{T_{s_{min}}}$, which measure the visibility of the X-ray pulses due to a `spot-like' glitch in one particular rotation around the peak of the luminosity-time graphs. The numerical values are given in Table~\ref{table:f}, where the magnetic field effect has not yet been included.

If $f_F\geq 0.05$, it should be large enough for observation, then the cases with $T_c=10^7\ K$, $\triangle E=10^{39}\ erg$ and $\rho _{glitch}=10^{11}\ g/cm^3$; $T_c=10^8\ K$, $\triangle E=10^{39}\ erg$ and $\rho _{glitch}=10^{10}\ g/cm^3$; $T_c=10^8\ K$, $\triangle E=10^{40}\ erg$ and $\rho _{glitch}=10^{11}\ g/cm^3$; and $T_c=10^7\ K$, $\triangle E=10^{42}\ erg$ and $\rho _{glitch}=10^{13}\ g/cm^3$ are possible to be observed.

In the legend of Fig.~\ref{fig:UTC8E42R13light418}, the first number is the angle of `spot' from the rotational axis and the second number is the angle of observer from the rotational axis. Other numerical results can be found in Tang (1999).

Taking the magnetic field effect into consideration, the background light curve is no longer a straight line. That is to say, the two magnetic poles have already produced modulation to the flux profiles. For easy calculation, the background light curves in the figures in this second part of the section are generated by using a $10^o\times 10^o$ grid; the rotational axis $\vec \Omega $ is taken to be the $\vec z$-axis; the magnetic moment $\vec \mu $ is taken to be at an angle of $45^o$ from $\vec \Omega $; the `spot' case glitch has its energy released at an angle of $60^o$ from $\vec \Omega $ and $90^o$ azimuthally from $\vec \mu $. In other words, if $(\vec \Omega ,\vec \mu )$ is the $z-y$ plane, then the spot is in the $z-x$ plane. The `ring' case glitch has its energy released along the rotational equator. The observer is at an angle of $45^o$ from $\vec \Omega $. 

For a `spot-like' glitch, the glitch effect alone can produce modulation to the X-ray profiles as mentioned in the beginning of this section. It can significantly increase the fraction $f_F$ with some particular sets of parameters. If there is a magnetic field effect, it can also cause a phase shift of the whole light curve as seen in Fig.~\ref{fig:spotC7B312E40R12} to Fig.~\ref{fig:spotC8B15E42R12}. The results are summarized in Table~\ref{table:fs}.

For the `ring' case, if there is no magnetic field, there will be no modulation to the light curve due to azimuthal symmetry. Only the observed total flux is increased. If a magnetic field effect is also considered, the glitch effect causes the peak of the X-ray profile to rise and the pulse shape to shift. However, it depresses the contrast $f_F$. This is easily observed from the values in Table~\ref{table:fr} and in Fig.~\ref{fig:ringC7B312E40R12} to Fig.~\ref{fig:ringC7B312E42R12}.

The rise of the peak flux value can be more drastic for the `spot' case than the `ring' case with the same set of parameters, especially at low temperature, e.g.\ $T_{core}=10^7\ K$, and with a large amount of energy deposition, e.g.\ $\triangle E=10^{42}\ erg$. The amount of phase shift of the light curve is determined by the position at which the glitch energy is released (its azimuthal angular distance from $\vec \mu $). For the `spot' case, since we choose the `spot' to be $90^o$ from $\vec \mu $ azimuthally, there is a $90^o$ phase shift in the X-ray light curve when the glitch energy reaches the surface (c.f.\ Fig.~\ref{fig:spotC7B312E42R12}). For the `ring' case, the additional peak results from the glitch (c.f. Fig.~\ref{fig:ringC7B312E42R12}). For both the `spot' and the `ring' cases, the values of $f_F$ are greater than $0.05$. As a result, we expect that the change of light curve due to the glitch should be observable. In fact, the change in pulse shapes point to the corresponding sets of parameters.

It is important to note that with the magnetic field effect, the heat transport becomes a 3-dimensional problem even for the `ring' case. The energy released in a glitch can have two routes. Part of it is transferred to the core (c.f. Fig.~\ref{fig:UTC7E42R13ringTr}). The core is able to quickly resume its isothermality and the energy is reradiated isotropically to both directions of the magnetic pole and the magnetic equator. According to Harding~and~Muslimov (1998), the transverse heat conductivity is suppressed due to magnetization of electrons, which results in a surface temperature at the magnetic pole higher than at the equator. The other part of the glitch energy is transferred to the crust, since the heat propagation time is proportional to the temperature squared as mentioned in Section~\ref{section:compare}. The distance from and the temperature of the heated area are the main elements affecting the angular distribution of the thermal energy on the surface.

When the `ring' case is compared to the `spot' case, the same amount of energy is deposited in a larger volume and the localized heating effect is averaged out. Therefore, the temperature in each particular volume element is not so high. Hence, the heat energy can reach the surface faster. Comparing Fig.~\ref{fig:UTC7E42R13ringTr}, which describes the thermal evolution of a `ring-like' glitch, to Fig.~1 in Cheng~and~Li (1997), which describes the thermal evolution of a `spot-like' glitch, this effect can be easily seen. That means, for a `ring-like' glitch, a larger part of the glitch energy is transferred to the crust than to the core. Since the magnetic pole and the magnetic equator are at the same distance from the `ring' (with the magnetic moment $\vec \mu $ is at an angle of $45^o$ from $\vec \Omega $ and the `ring' is at $90^o$ from $\vec \Omega $), and the magnetic pole is hotter, energy transport in the magnetic equator direction is faster hence decreasing the contrast $f_F$ as shown in Table~\ref{table:fr}. For the `spot' case, however, a larger fraction of the glitch energy is transported to the core, which explains the increase in the contrast $f_F$ for the `spot' case.

For the case of low temperature, $T_{core}=10^7\ K$, and large energy deposition, $\triangle E=10^{42}\ erg$, since the localized heating is enormous, the effect of polar cap heating can be viewed simply as a modification to the uniform temperature profile. The magnetic effect is absolutely overwhelmed by the glitch effect. However, for a young star like Vela, the core temperature is still high ($\sim 10^8\ K$), the rate of energy transport is slower, and hence a larger fraction of the glitch energy is transferred to the core. Therefore, the glitch effect is not significant both in terms of phase shift and the rise of the peak of flux. 

We define a quantity $E_\gamma =\int _0^{2t_{max}}(L(t)-L_{BG})dt$, where $t_{max}$ is the time at which $L(t)$ reaches its peak value after the glitch and $L_{BG}$ is the luminosity before the glitch. This parameter indicates how much glitch energy is released as a pulse. From the "$E_\gamma $" column in Table~\ref{table:fs} and Table~\ref{table:fr}, it can be easily seen that the faster the luminosity ($L$) reaches its peak value, the smaller is the fraction of the glitch energy eventually emitted as photons at the surface within $2t_{max}$. We wish to point out that eventually all the glitch energy will be radiated through the surface in a time scale much longer than $t_{max}$.

\section{Conclusion and discussion \label{section:conclusion}}
In this paper, we have employed the relativistic heat diffusion equations to calculate the thermal response of a pulsar to a glitch with energy deposited in a ring-like area around the rotational equator in the inner crust, in a spherical shell and in a small volume. The `ring' case is always the middle one in terms of the response time, the response period and the intensity of the response. This is not unexpected as the volume of energy deposition is smaller than the `shell' case but larger than the `spot' case. Moreover, we have calculated the expected light curves for the thermal X-ray regime as observed after a glitch for the `spot' case and the `ring' case. The `shell' case is ignored because we believe that the modulation to the light curves in this case will be too insignificant to be observed. After a glitch, the energy flux emitted from the surface is not uniform. Excess thermal X-rays are emitted at a particular phase. This alters the X-ray pulse shape of the pulsar. These curves can be used to compare with the actual data obtained to put constraints on the true glitch model as well as the interior structure of a neutron star.

For the prompt afterglow of a glitch to be detectable, the energy released should be enormous so that it should re-radiate over a reasonable time scale and the change in intensity of the radiation can be observed. Thermal emissions for $\triangle E\gtrsim 10^{42}\ erg$ lie in the soft X-ray region (Chong~and~Cheng 1994). The thermal transients produced by giant glitches in nearby pulsars may be observable by Chandra, XMM and Astro-E. These three missions have complementary capabilities in measuring soft X-ray pulses from neutron stars.

For observations taken following a glitch that triggers thermal afterglow, one should be able to detect significant changes in the X-ray pulse shape if the energy release is anisotropic. The results will be valuable in refining neutron star equations of state and gaining a better understanding of the physics behind neutron star transient energy releases (Anderson~and~\"Ogelman 1997).

The defect of this method of finding the internal properties of a neutron star is that if the glitch events occur too frequently, it may result in a pile-up of the pulses as well as a long-term variation of the total thermal radiation which eventually reduces the detectability of the thermal afterglow of the glitch (Li  1997). In the calculations in this paper, several assumptions have been made: the interstellar absorption, the magnetospheric effects and the possibility of other diffusion mechanisms besides conduction in the pulsar have been neglected; the structure of the pulsar is assumed to be inert to the thermal change of the pulsar; the thermal X-ray emitted from the surface is assumed to be blackbody radiation alone. In future study, we shall take these factors into account.

\acknowledgments{We thank J.S. Heyl, D. Page and S. Tsuruta for useful comments and discussion and Terry Boyce for a critical reading. This work is partially supported by a RGC grant from the Hong Kong Government and a Croucher Foundation Senior Research Fellowship.}

\clearpage  
\begin{table}[h]
\caption{The fractions $f_F$ and $f_T$ are calculated for various sets of parameters with the `spot' case of UT equation of state with no $\vec B$ field and spot and the observer both lying along the rotational equator.\label{table:f}} 
\begin{center}
\begin{tabular}{|ccc|r|r|}
\hline
$T_{core}\ (K)$ & $\triangle E\ (erg)$ & $\rho _{glitch}\ (g/cm^3)$ & $f_F$ & $f_T$ \\ \hline
$10^8$ & $10^{38}$ & $10^{11}$ & $0.0002$ & $0.0395$ \\
$10^8$ & $10^{39}$ & $10^{11}$ & $0.0033$ & $0.3400$ \\
$10^8$ & $10^{40}$ & $10^{11}$ & $0.0517$ & $1.4493$ \\
$10^8$ & $10^{41}$ & $10^{11}$ & $1.8673$ & $4.9635$ \\
$10^8$ & $10^{42}$ & $10^{11}$ & $50.8427$ & $12.6201$ \\ \hline
$10^8$ & $10^{38}$ & $10^{10}$ & $0.0033$ & $0.3411$ \\
$10^8$ & $10^{39}$ & $10^{10}$ & $0.0814$ & $1.7365$ \\
$10^8$ & $10^{40}$ & $3\times 10^{11}$ & $0.0097$ & $0.6568$ \\
$10^8$ & $10^{42}$ & $10^{13}$ & $0.0776$ & $1.7047$ \\ \hline
$10^7$ & $10^{38}$ & $10^{11}$ & $0.0183$ & $0.9121$ \\
$10^7$ & $10^{39}$ & $10^{11}$ & $0.3670$ & $3.8872$ \\
$10^7$ & $10^{42}$ & $10^{12}$ & $66.2741$ & $13.5537$ \\
$10^7$ & $10^{42}$ & $10^{13}$ & $9.0100$ & $7.8376$ \\ \hline
\end{tabular} 
\end{center}
\end{table}

\begin{table}[h]
\caption{The fractions $f_F$ are calculated for various sets of parameters for the `spot' case of UT equation of state with $\vec B$ at $45^o$ from $\vec \Omega $, the `spot' located at an angle of $60^o$ from $\vec \Omega $ and $90^o$ azimuthally from $\vec B$, and the observer at an angle $45^o$ with $\vec \Omega $. The `new peak' represents the maximum flux value after the glitch and the `old peak' represents the value before the glitch. The value $E_\gamma =\int _0^{2t_{max}}(L(t)-L_{BG})dt$ is the amount of energy that is emitted as photons at the surface. $t_{max}$ is the time at which the luminosity has its peak value after the glitch. $L_{BG}$ is the background luminosity.\label{table:fs}} 
\begin{center}
\begin{tabular}{|ccc|c|c|c|c|c|}
\hline
$T_{core}$ & $\triangle E$ & $\rho _{glitch}$ & $E_\gamma $ & $\vec B$ & $f_F$ & $f_F$ & new peak\\
$(K)$ & $(erg)$ & $(g/cm^3)$ & $(erg)$ & $(G)$ & before glitch & at $t_{max}$ & to old peak\\ \hline
 & & & & $0$ & $0.00$ & $\ 0.09$ & $\ 1.09$\\
$10^7$ & $10^{40}$ & $10^{12}$ & $3.97\times 10^{35}$ & $3\times 10^{12}$ & $0.41$ & $\ 0.42$ & $\ 1.02$\\
 & & & & $10^{15}$ & $0.41$ & $\ 0.41$ & $\ 1.00$\\ \hline
 & & & & $0$ & $0.00$ & $\ 0.00$ & $\ 1.00$\\
$10^8$ & $10^{40}$ & $10^{12}$ & $5.07\times 10^{37}$ & $3\times 10^{12}$ & $0.41$ & $\ 0.41$ & $\ 1.00$\\
 & & & & $10^{15}$ & $0.40$ & $\ 0.40$ & $\ 1.00$\\ \hline
 & & & & $0$ & $0.00$ & $14.52$ & $56.76$\\
$10^7$ & $10^{42}$ & $10^{12}$ & $1.91\times 10^{39}$ & $3\times 10^{12}$ & $0.41$ & $13.20$ & $31.86$\\
 & & & & $10^{15}$ & $0.41$ & $13.23$ & $31.78$\\ \hline
 & & & & $0$ & $0.00$ & $\ 0.34$ & $\ 1.36$\\
$10^8$ & $10^{42}$ & $10^{12}$ & $1.91\times 10^{39}$ & $3\times 10^{12}$ & $0.41$ & $\ 0.49$ & $\ 1.10$\\
 & & & & $10^{15}$ & $0.40$ & $\ 0.49$ & $\ 1.10$\\ \hline
\end{tabular} 
\end{center}
\end{table}

\begin{table}[h]
\caption{The fractions $f_F$ are calculated for various sets of parameters for the `ring' case of UT equation of state with $\vec \mu $ as well as the observer making an angle $45^o$ with $\vec \Omega $. The `new peak' represents the maximum flux value after the glitch and the `old peak' represents this value before the glitch. The value $E_\gamma =\int _0^{2t_{max}}(L(t)-L_{BG})dt$ is the amount of energy that is emitted as photons at the surface. $t_{max}$ is the time at which the luminosity has its peak value after the glitch. $L_{BG}$ is the background luminosity.\label{table:fr}} 
\begin{center}
\begin{tabular}{|ccc|c|c|c|c|c|}
\hline
$T_{core}$ & $\triangle E$ & $\rho _{glitch}$ & $E_\gamma $ & $\vec B$ & $f_F$ &  $f_F$ & new peak\\
$(K)$ & $(erg)$ & $(g/cm^3)$ & $(erg)$ & $(G)$ & before glitch & at $t_{max}$ & to old peak\\ \hline
 & & & & $0$ & $0.00$ & $0.00$ & $\ 1.09$\\
$10^7$ & $10^{40}$ & $10^{12}$ & $2.00\times 10^{35}$ & $3\times 10^{12}$ & $0.41$ & $0.38$ & $\ 1.07$\\
 & & & & $10^{15}$ & $0.41$ & $0.38$ & $\ 1.07$\\ \hline
 & & & & $0$ & $0.00$ & $0.00$ & $\ 1.00$\\
$10^8$ & $10^{40}$ & $10^{12}$ & $1.68\times 10^{35}$ & $3\times 10^{12}$ & $0.40$ & $0.41$ & $\ 1.00$\\
 & & & & $10^{15}$ & $0.41$ & $0.40$ & $\ 1.00$\\ \hline
 & & & & $0$ & $0.00$ & $0.00$ & $11.61$\\
$10^7$ & $10^{42}$ & $10^{12}$ & $9.80\times 10^{37}$ & $3\times 10^{12}$ & $0.41$ & $0.20$ & $\ 9.66$\\
 & & & & $10^{15}$ & $0.41$ & $0.20$ & $\ 9.65$\\ \hline
 & & & & $0$ & $0.00$ & $0.00$ & $\ 1.07$\\
$10^8$ & $10^{42}$ & $10^{12}$ & $1.02\times 10^{38}$ & $3\times 10^{12}$ & $0.40$ & $0.38$ & $\ 1.06$\\
 & & & & $10^{15}$ & $0.41$ & $0.38$ & $\ 1.06$\\ \hline
\end{tabular} 
\end{center}
\end{table}

\clearpage
\begin{center}
\bf{ Figure Captions}
\end{center}

\figcaption{The redshifted temperature profile of a UT-star with $T_c=10^7\ K$, $\triangle E=10^{42}\ erg$ at $\rho _{glitch}=10^{13}\ g/cm^3$ for the `ring' case with respect to the radius from the centre of the star. The top left panel indicates the number of days after the glitch. \label{fig:UTC7E42R13ringTr}}

\figcaption{The redshifted surface temperature of a UT-star with $T_c=10^7\ K$, $\triangle E=10^{42}\ erg$ at $\rho _{glitch}=10^{13}\ g/cm^3$ for the `ring' case with respect to the angle from the rotational axis. \label{fig:UTC7E42R13ringTa}}

\figcaption{The redshifted surface temperature of a UT-star with $T_c=10^7\ K$, $\triangle E=10^{42}\ erg$ at $\rho _{glitch}=10^{13}\ g/cm^3$ for the `spot' case with respect to the angle from the rotational axis. The numbers besides the lines are the number of days after the glitch. \label{fig:UTC7E42R13spotTa}}

\figcaption{The luminosity of a UT-star with $T_c=10^7\ K$, $\triangle E=10^{42}\ erg$ at $\rho _{glitch}=10^{13}\ g/cm^3$. The three cases-`shell', `ring' and `spot'-are compared. \label{fig:UTC7E42R13}}

\figcaption{The luminosity of a UT-star with $T_c=10^8\ K$, $\triangle E=10^{42}\ erg$ at $\rho _{glitch}=10^{13}\ g/cm^3$. The three cases-`shell', `ring' and `spot'-are compared. \label{fig:UTC8E42R13}}

\figcaption{The luminosity profiles of a BPS-star, a UT-star and a PPS-star with $T_c=10^7\ K$, $\triangle E=10^{42}\ erg$ at $\rho _{glitch}=10^{13}\ g/cm^3$ for the `ring' case. \label{fig:C7E42R13ringEOS}}

\figcaption{Schematic illustration of gravitational lensing effect. Angles are defined as shown.\label{fig:delta}}

\figcaption{Gravitational lensing effect for different EOS.\label{fig:angle}}

\figcaption{Angles used in calculating the rotational effect.\label{fig:theta0}}

\figcaption{Finite area for the `spot' is considered for different EOS: BPS, PPS and UT. The angular radius of the emission cone is $\alpha =5^o$. \label{fig:brightEOS}}

\figcaption{The evolution of the total brightness for a UT-star with $T_c=10^8\ K$ at $\rho _{glitch}=10^{11}\ g/cm^3$ with $\triangle E=10^{40}\ erg$ for the `spot' case. \label{fig:UTC8E40R11chp5f}}

\figcaption{The total brightness for a UT-star with $T_c=10^8\ K$ at $\rho _{glitch}=10^{13}\ g/cm^3$ with $\triangle E=10^{42}\ erg$ 418 days after a `spot-like' glitch. \label{fig:UTC8E42R13light418}}

\figcaption{The evolution of the total brightness for a UT-star with $T_c=10^7\ K$ at $\rho _{glitch}=10^{12}\ g/cm^3$ with $\triangle E=10^{40}\ erg$ for the `spot' case with magnetic field $\vec B=3\times 10^{12}\ G$. \label{fig:spotC7B312E40R12}}

\figcaption{The evolution of the total brightness for a UT-star with $T_c=10^7\ K$ at $\rho _{glitch}=10^{12}\ g/cm^3$ with $\triangle E=10^{42}\ erg$ for the `spot' case with magnetic field $\vec B=3\times 10^{12}\ G$. \label{fig:spotC7B312E42R12}}

\figcaption{The evolution of the total brightness for a UT-star with $T_c=10^8\ K$ at $\rho _{glitch}=10^{12}\ g/cm^3$ with $\triangle E=10^{40}\ erg$ for the `spot' case with magnetic field $\vec B=3\times 10^{12}\ G$. \label{fig:spotC8B312E40R12}}

\figcaption{The evolution of the total brightness for a UT-star with $T_c=10^8\ K$ at $\rho _{glitch}=10^{12}\ g/cm^3$ with $\triangle E=10^{42}\ erg$ for the `spot' case with magnetic field $\vec B=3\times 10^{12}\ G$. \label{fig:spotC8B312E42R12}}

%\figcaption{The evolution of the total brightness for a UT-star with $T_c=10^7\ K$ at $\rho _{glitch}=10^{12}\ g/cm^3$ with $\triangle %E=10^{40}\ erg$ for the `spot' case with magnetic field $\vec B=10^{15}\ G$. \label{fig:spotC7B15E40R12}}

%\figcaption{The evolution of the total brightness for a UT-star with $T_c=10^7\ K$ at $\rho _{glitch}=10^{12}\ g/cm^3$ with $\triangle %E=10^{42}\ erg$ for the `spot' case with magnetic field $\vec B=10^{15}\ G$. \label{fig:spotC7B15E42R12}}

%\figcaption{The evolution of the total brightness for a UT-star with $T_c=10^8\ K$ at $\rho _{glitch}=10^{12}\ g/cm^3$ with $\triangle %E=10^{40}\ erg$ for the `spot' case with magnetic field $\vec B=10^{15}\ G$. \label{fig:spotC8B15E40R12}}

\figcaption{The evolution of the total brightness for a UT-star with $T_c=10^8\ K$ at $\rho _{glitch}=10^{12}\ g/cm^3$ with $\triangle E=10^{42}\ erg$ for the `spot' case with magnetic field $\vec B=10^{15}\ G$. \label{fig:spotC8B15E42R12}}

\figcaption{The evolution of the total brightness for a UT-star with $T_c=10^7\ K$ at $\rho _{glitch}=10^{12}\ g/cm^3$ with $\triangle E=10^{40}\ erg$ for the `ring' case with magnetic field $\vec B=3\times 10^{12}\ G$. \label{fig:ringC7B312E40R12}}

\figcaption{The evolution of the total brightness for a UT-star with $T_c=10^7\ K$ at $\rho _{glitch}=10^{12}\ g/cm^3$ with $\triangle E=10^{42}\ erg$ for the `ring' case with magnetic field $\vec B=3\times 10^{12}\ G$. \label{fig:ringC7B312E42R12}}

%\figcaption{The evolution of the total brightness for a UT-star with $T_c=10^8\ K$ at $\rho _{glitch}=10^{12}\ g/cm^3$ with $\triangle %E=10^{40}\ erg$ for the `ring' case with magnetic field $\vec B=3\times 10^{12}\ G$. \label{fig:ringC8B312E40R12}}

\figcaption{The evolution of the total brightness for a UT-star with $T_c=10^8\ K$ at $\rho _{glitch}=10^{12}\ g/cm^3$ with $\triangle E=10^{42}\ erg$ for the `ring' case with magnetic field $\vec B=3\times 10^{12}\ G$. \label{fig:ringC8B312E42R12}}

%\figcaption{The evolution of the total brightness for a UT-star with $T_c=10^7\ K$ at $\rho _{glitch}=10^{12}\ g/cm^3$ with $\triangle %E=10^{40}\ erg$ for the `ring' case with magnetic field $\vec B=10^{15}\ G$. \label{fig:ringC7B15E40R12}}

%\figcaption{The evolution of the total brightness for a UT-star with $T_c=10^7\ K$ at $\rho _{glitch}=10^{12}\ g/cm^3$ with $\triangle %E=10^{42}\ erg$ for the `ring' case with magnetic field $\vec B=10^{15}\ G$. \label{fig:ringC7B15E42R12}}

%\figcaption{The evolution of the total brightness for a UT-star with $T_c=10^8\ K$ at $\rho _{glitch}=10^{12}\ g/cm^3$ with $\triangle %E=10^{40}\ erg$ for the `ring' case with magnetic field $\vec B=10^{15}\ G$. \label{fig:ringC8B15E40R12}}

%\figcaption{The evolution of the total brightness for a UT-star with $T_c=10^8\ K$ at $\rho _{glitch}=10^{12}\ g/cm^3$ with $\triangle %E=10^{42}\ erg$ for the `ring' case with magnetic field $\vec B=10^{15}\ G$. \label{fig:ringC8B15E42R12}}

\begin{figure}[ht]
\vbox to7.2in{\rule{0pt}{7.2in}}
\includegraphics{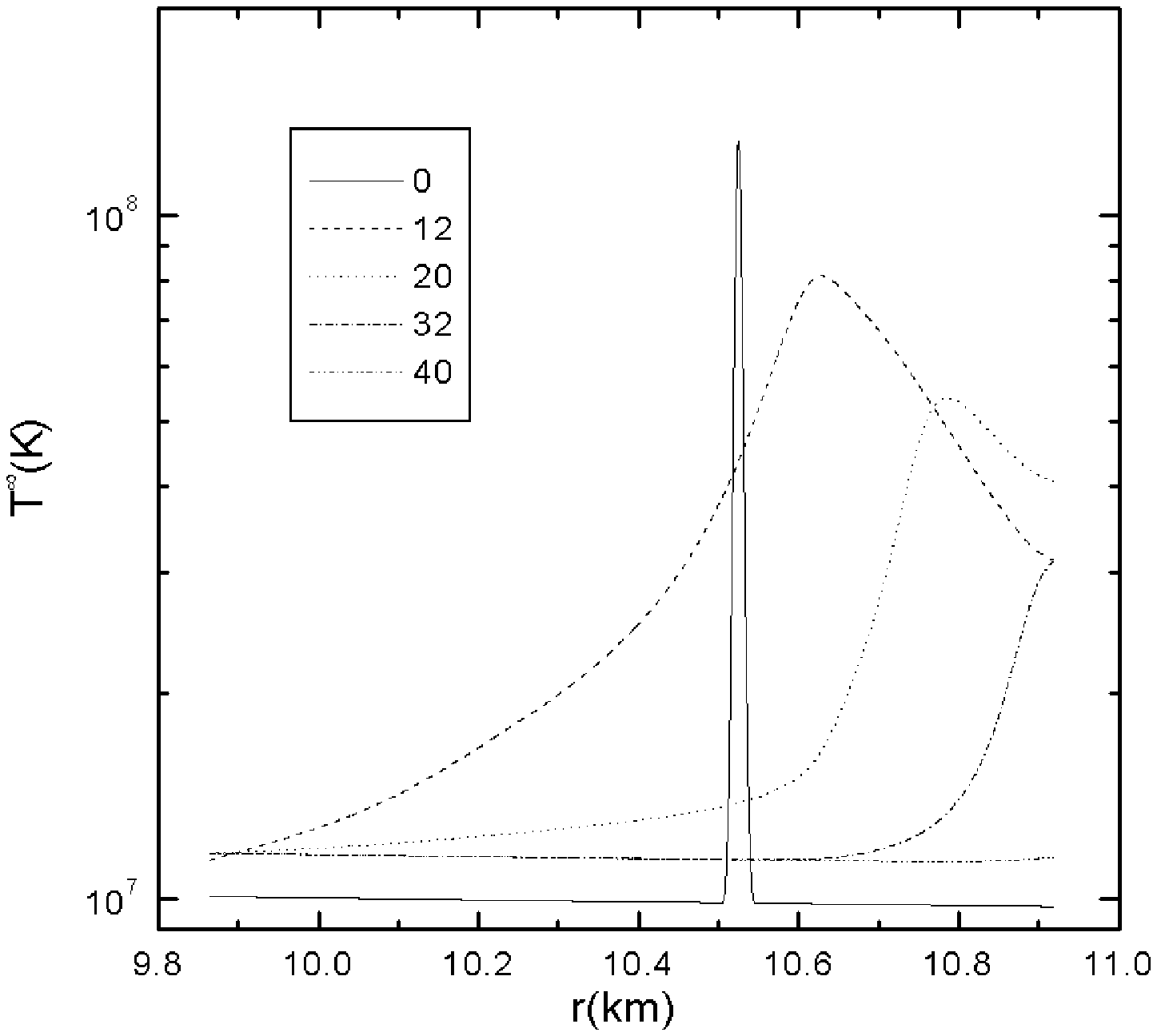}
\noindent{Fig.~\ref{fig:UTC7E42R13ringTr}}
\end{figure}

\begin{figure}[ht]
\vbox to7.2in{\rule{0pt}{7.2in}}
\includegraphics{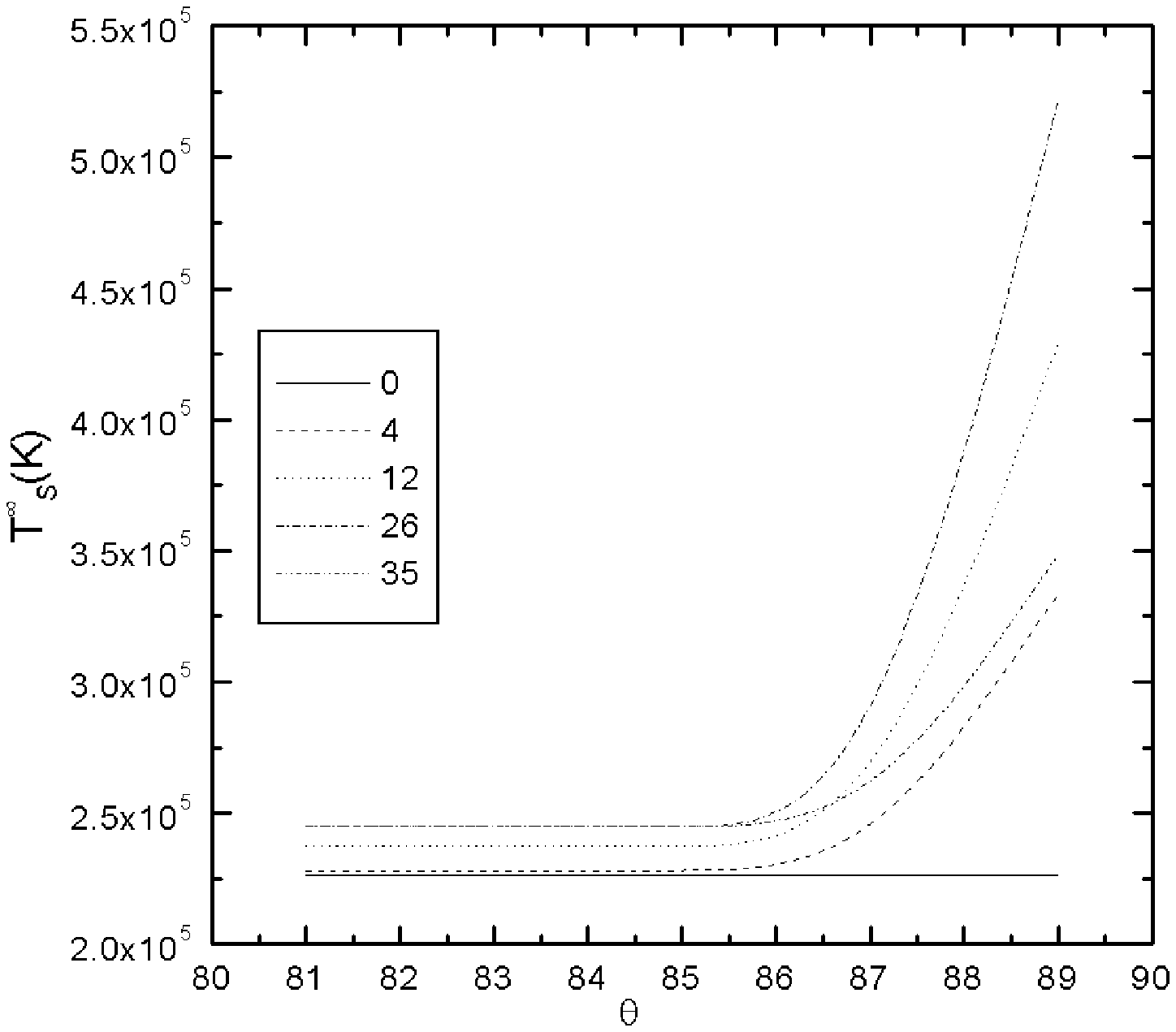}
\noindent{Fig.~\ref{fig:UTC7E42R13ringTa}}
\end{figure}

\begin{figure}[ht]
\vbox to7.2in{\rule{0pt}{7.2in}}
\includegraphics{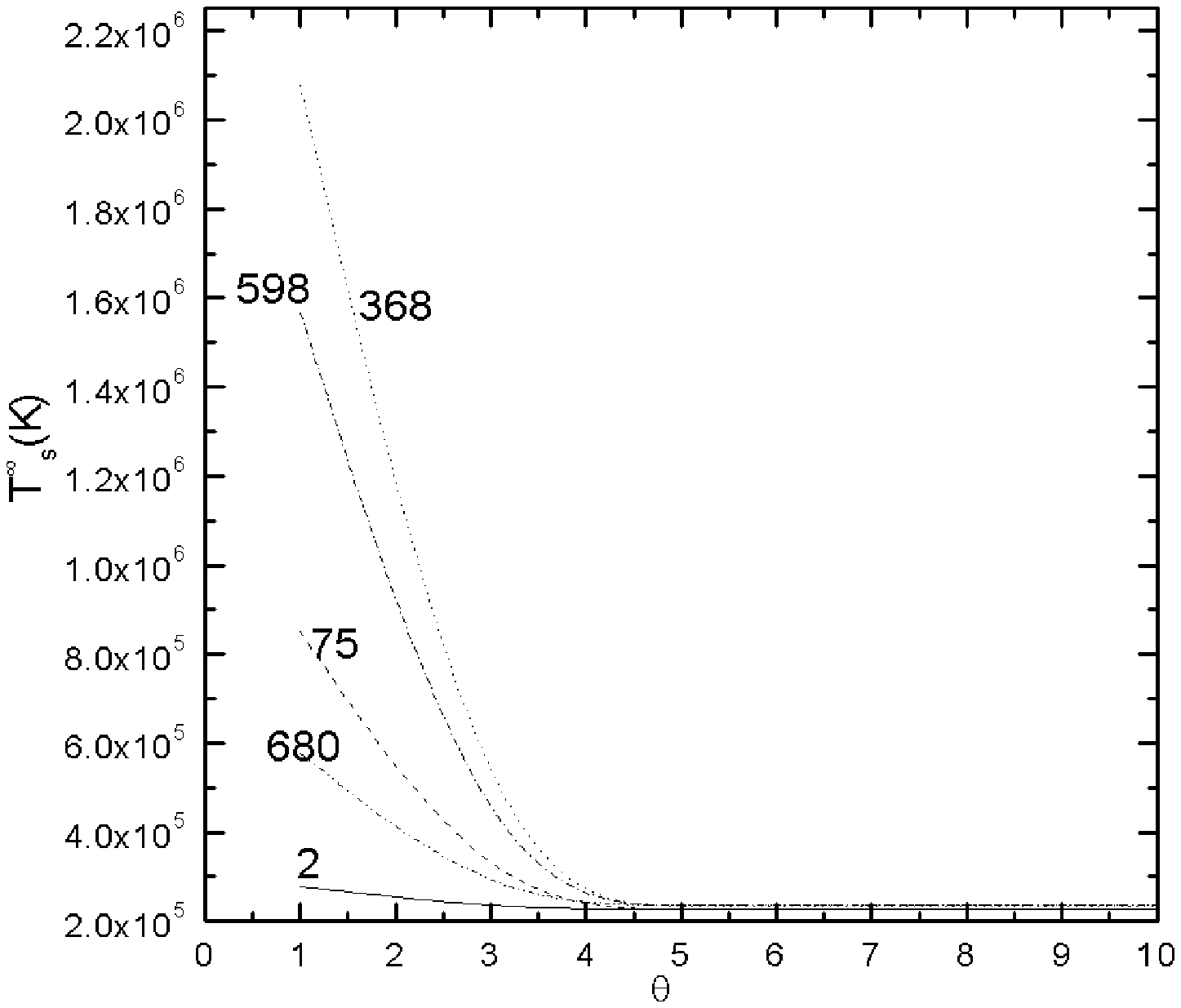}
\noindent{Fig.~\ref{fig:UTC7E42R13spotTa}}
\end{figure}

\begin{figure}[ht]
\vbox to7.2in{\rule{0pt}{7.2in}}
\includegraphics{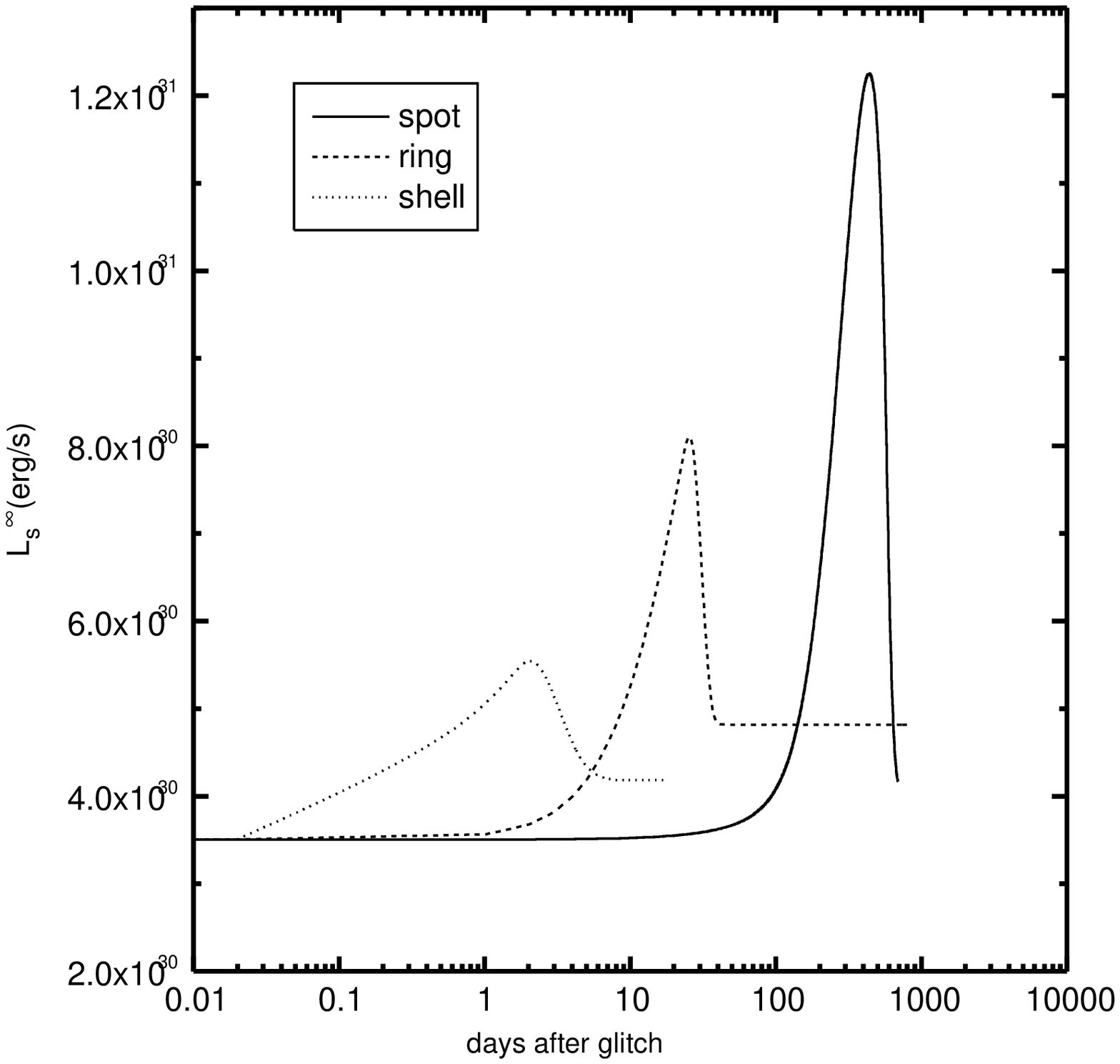}
\noindent{Fig.~\ref{fig:UTC7E42R13}}
\end{figure}

\begin{figure}[ht]
\vbox to7.2in{\rule{0pt}{7.2in}}
\includegraphics{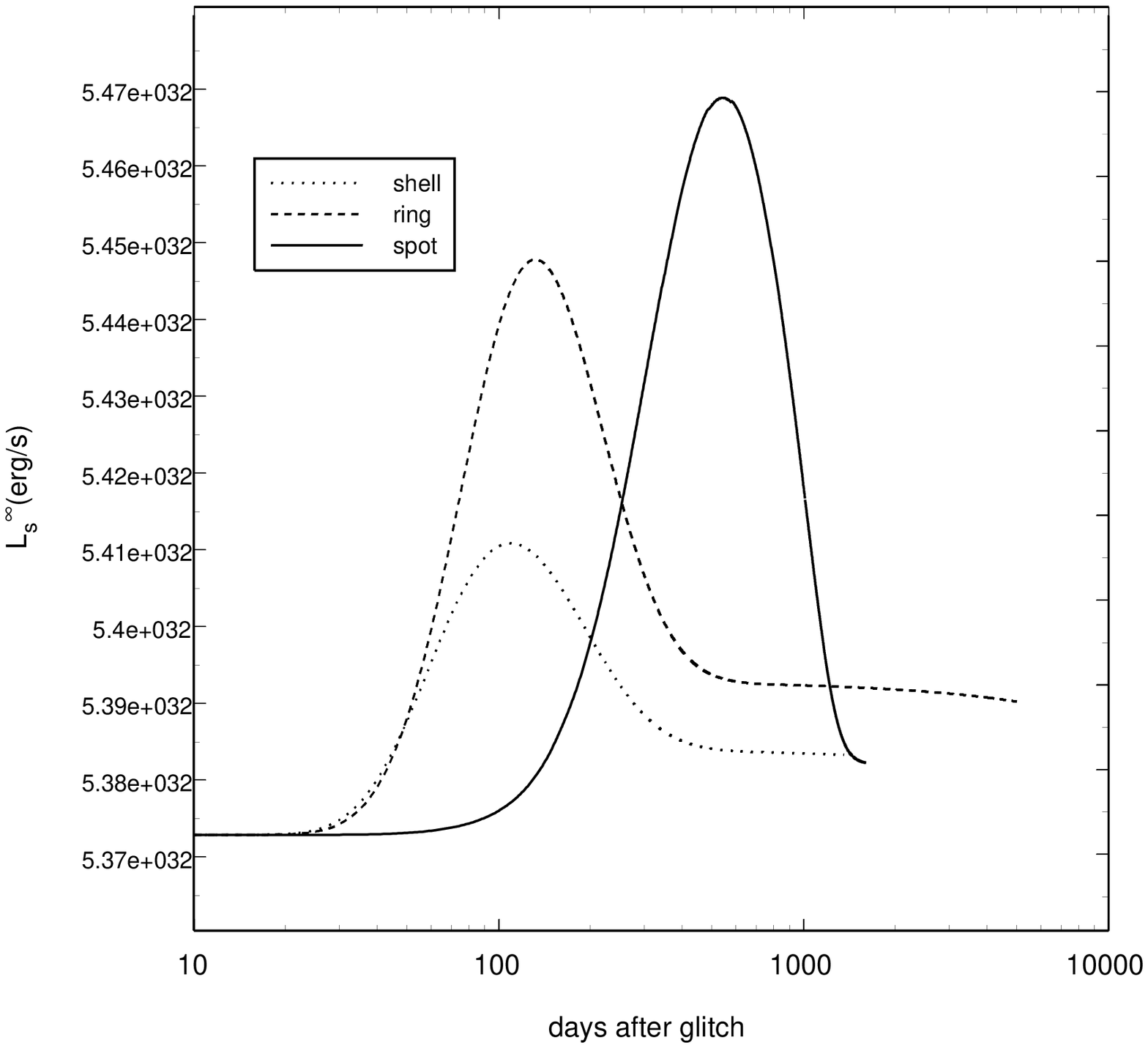}
\noindent{Fig.~\ref{fig:UTC8E42R13}}
\end{figure}

\begin{figure}[ht]
\vbox to7.2in{\rule{0pt}{7.2in}}
\includegraphics{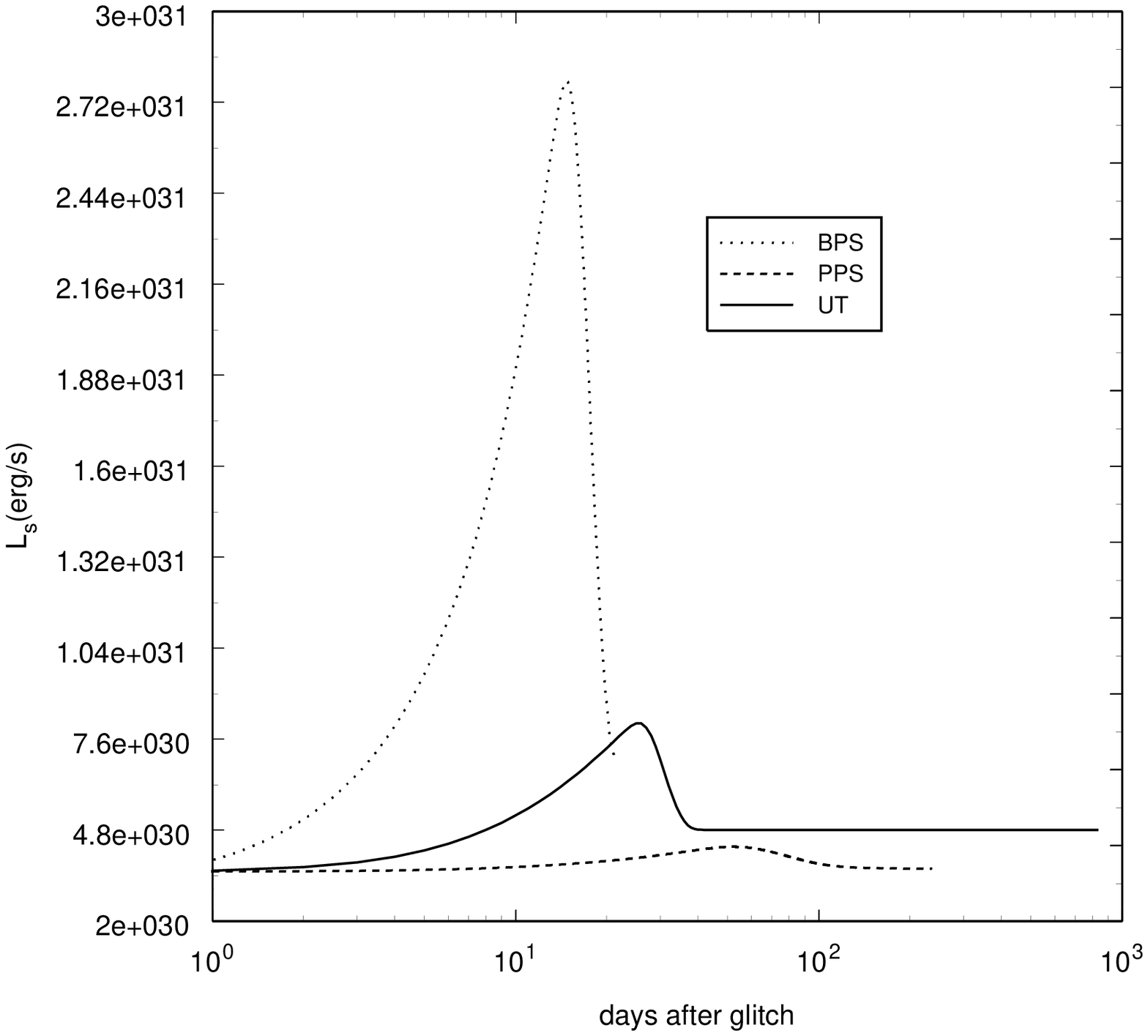}
\noindent{Fig.~\ref{fig:C7E42R13ringEOS}}
\end{figure}

\begin{figure}[ht]
\vbox to7.2in{\rule{0pt}{7.2in}}
\includegraphics{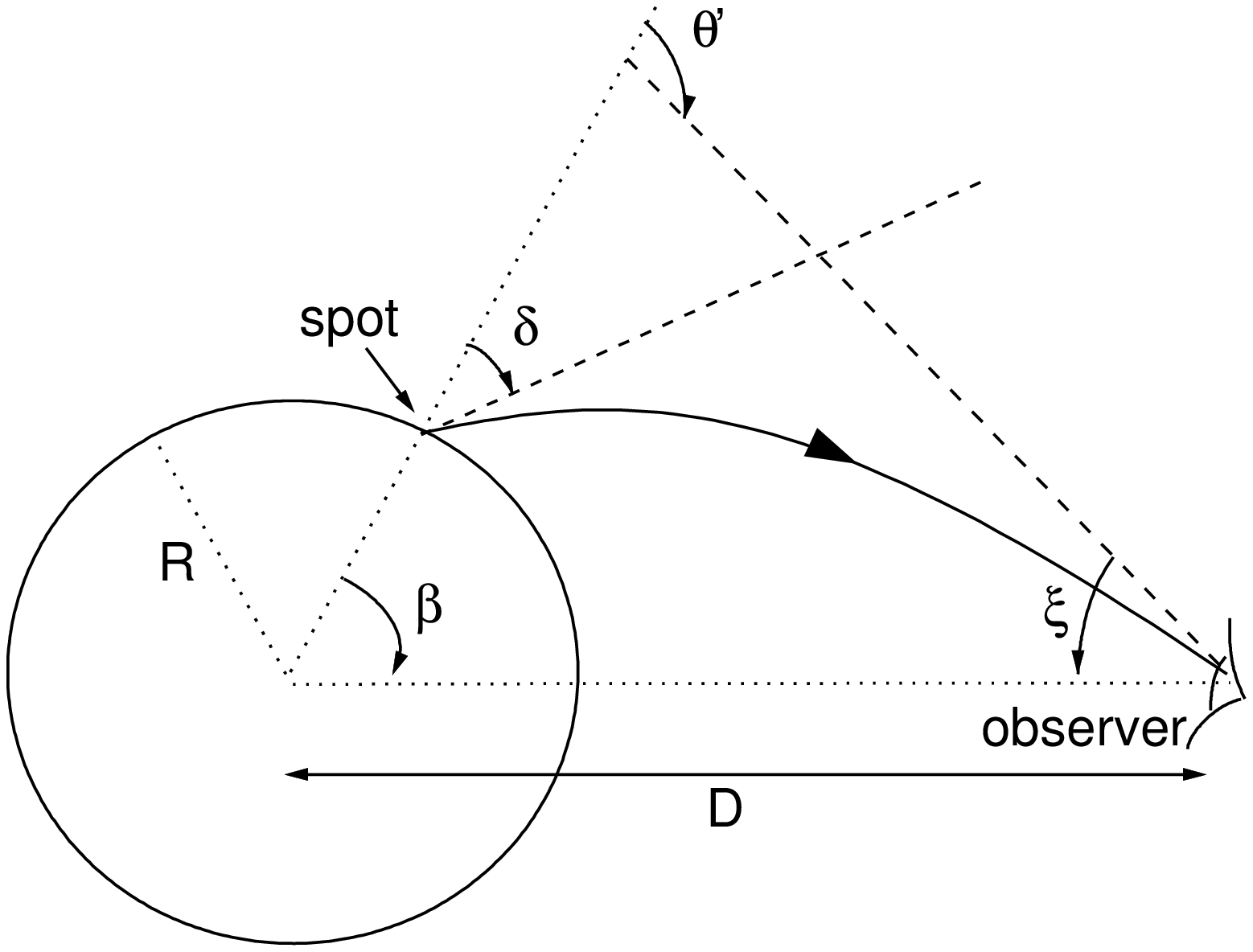} 
\noindent{Fig.~\ref{fig:delta}}
\end{figure}

\begin{figure}[ht]
\vbox to7.2in{\rule{0pt}{7.2in}}
\includegraphics{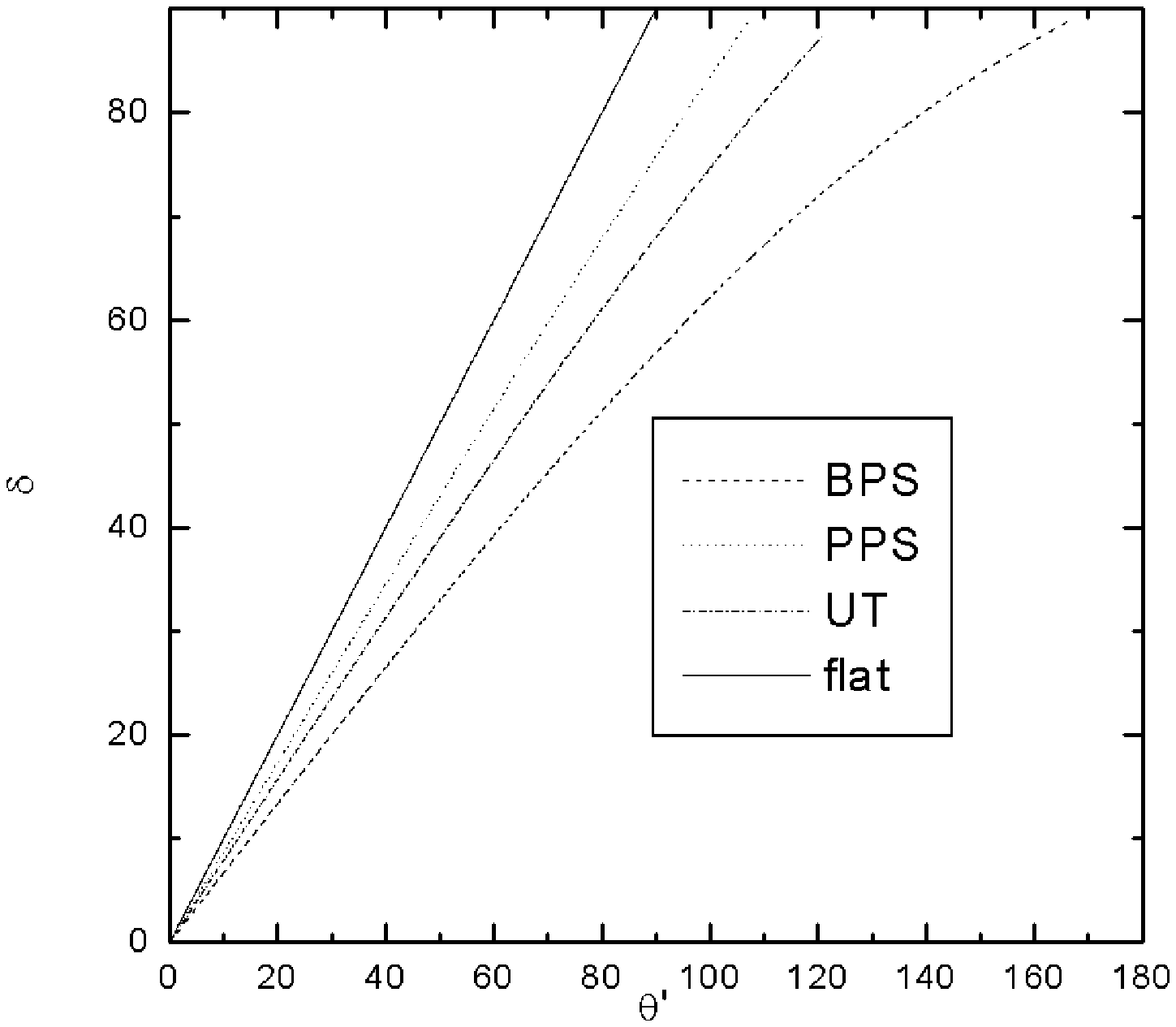}
\noindent{Fig.~\ref{fig:angle}}
\end{figure}

\begin{figure}[ht]
\vbox to7.2in{\rule{0pt}{7.2in}}
\includegraphics{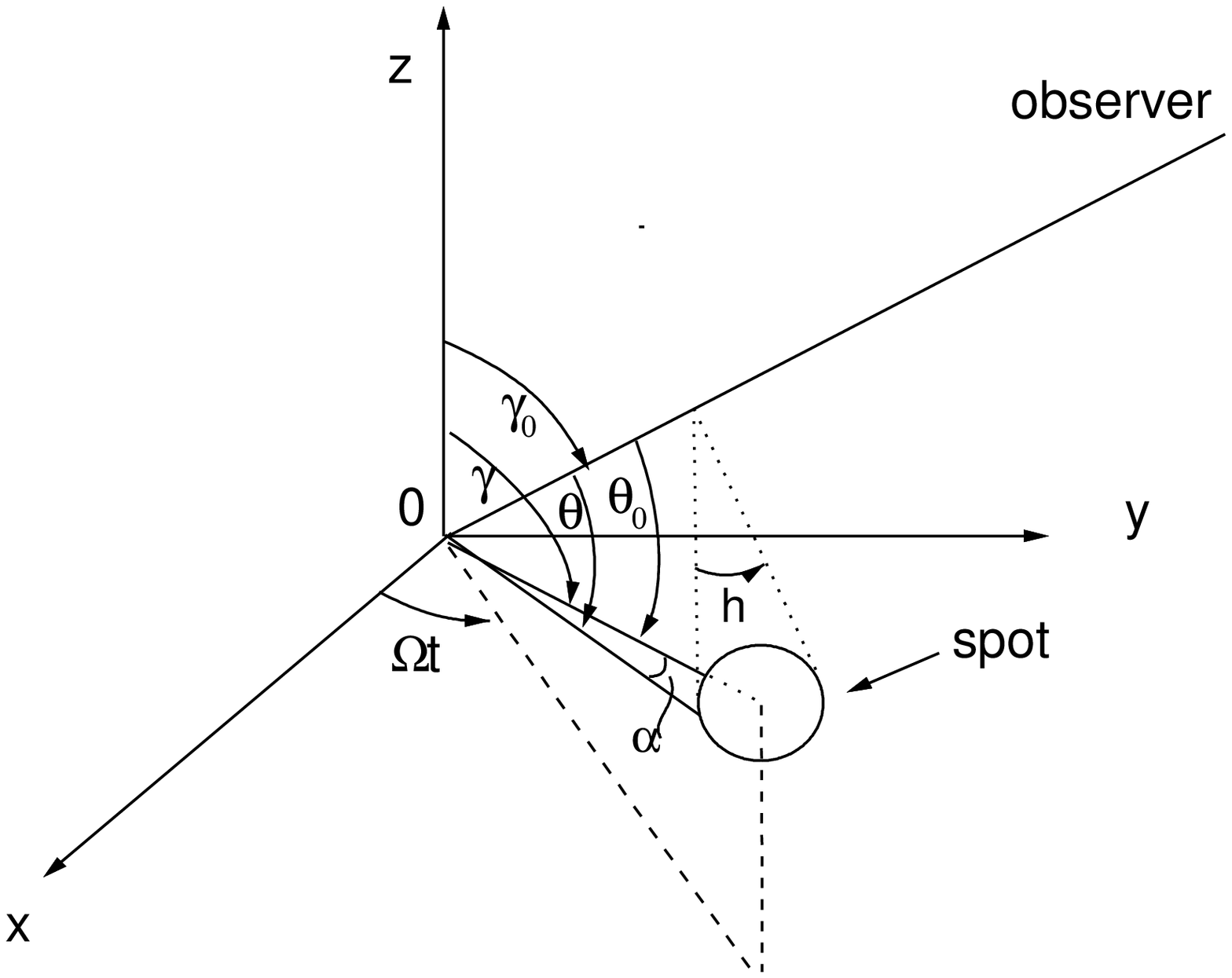}
\noindent{Fig.~\ref{fig:theta0}}
\end{figure}

\begin{figure}[ht]
\vbox to7.2in{\rule{0pt}{7.2in}}
\includegraphics{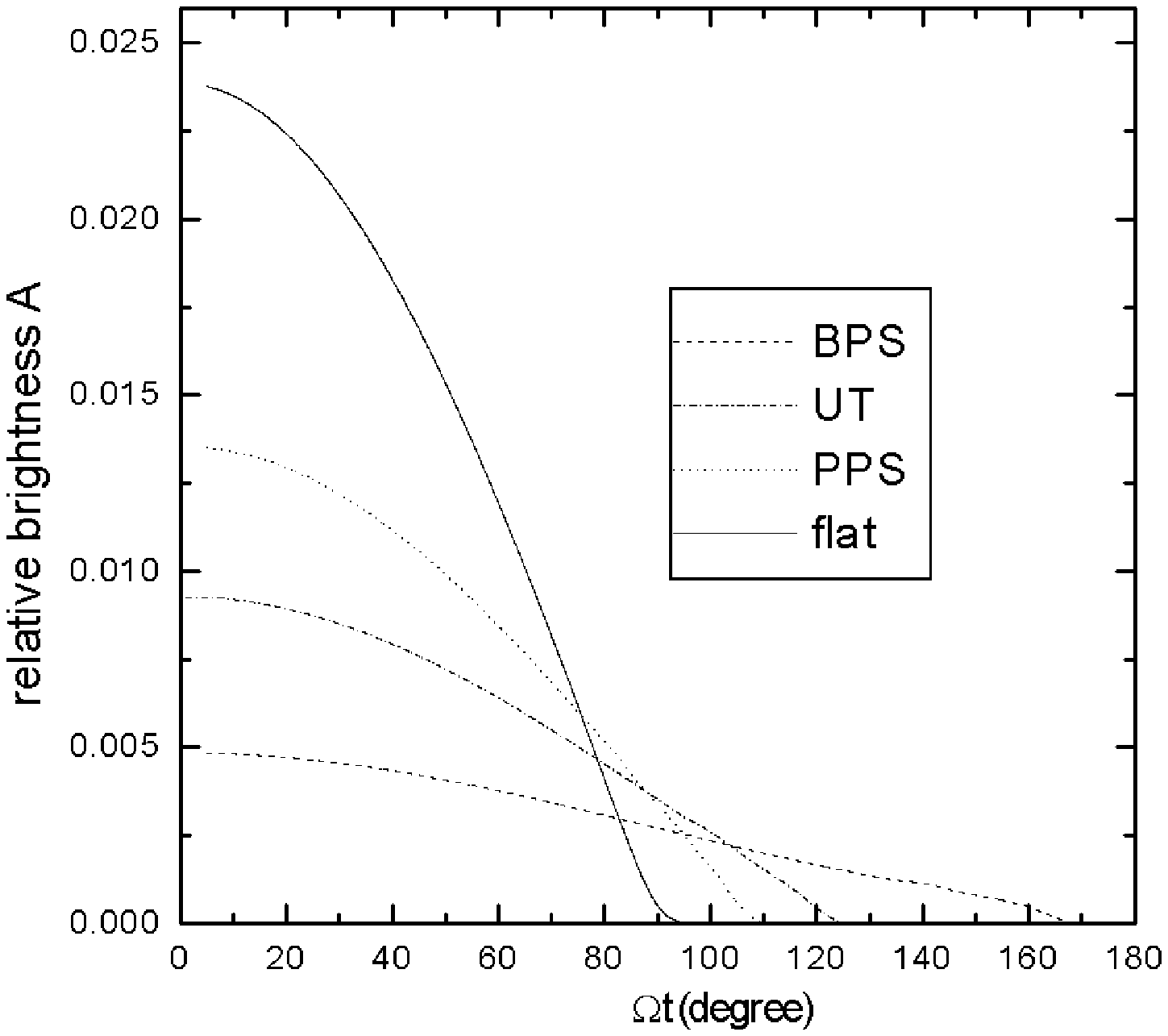}
\noindent{Fig.~\ref{fig:brightEOS}}
\end{figure}

\begin{figure}[ht]
\vbox to7.2in{\rule{0pt}{7.2in}}
\includegraphics{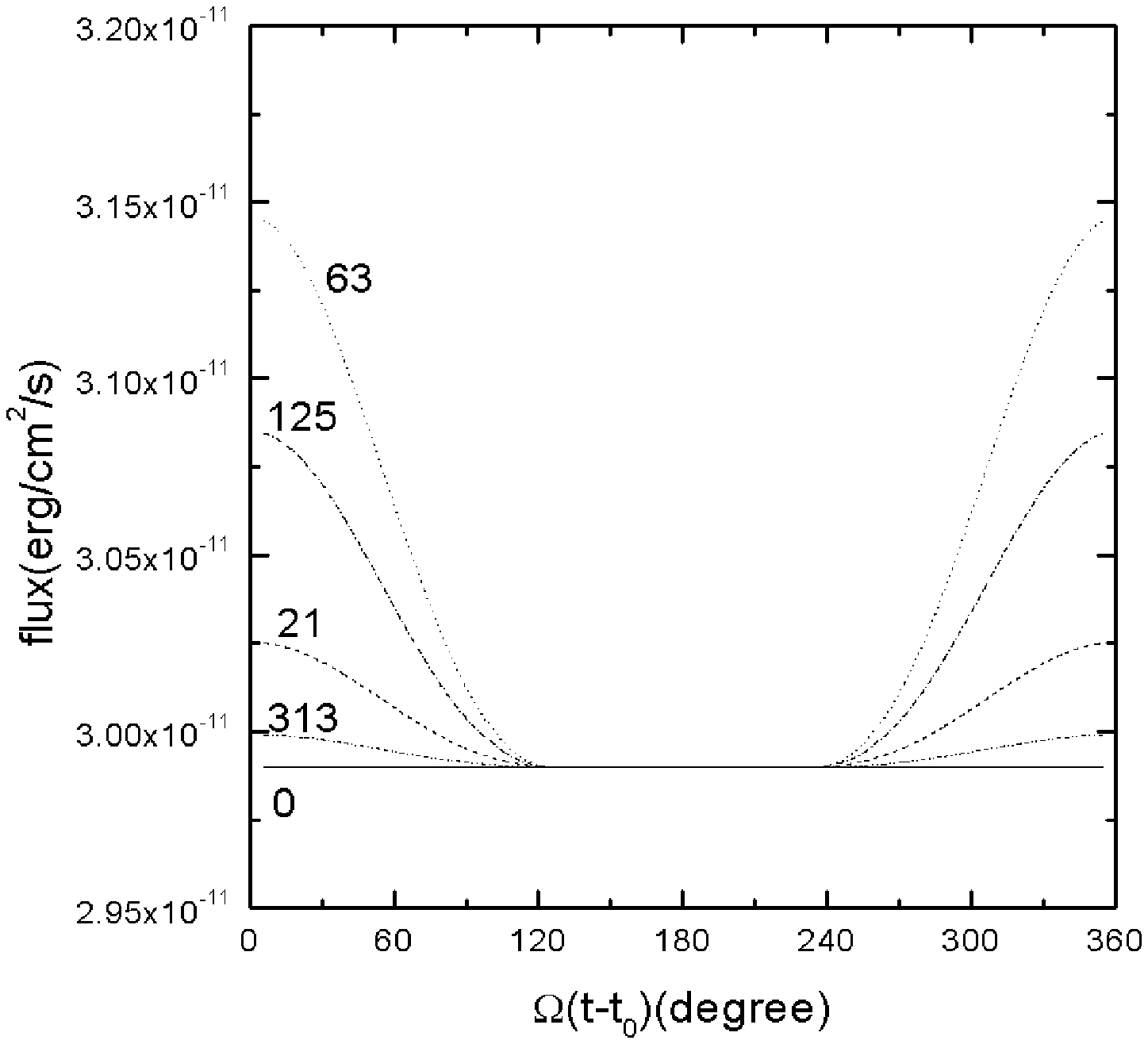}
\noindent{Fig.~\ref{fig:UTC8E40R11chp5f}}
\end{figure}

\begin{figure}[ht]
\vbox to7.2in{\rule{0pt}{7.2in}}
\includegraphics{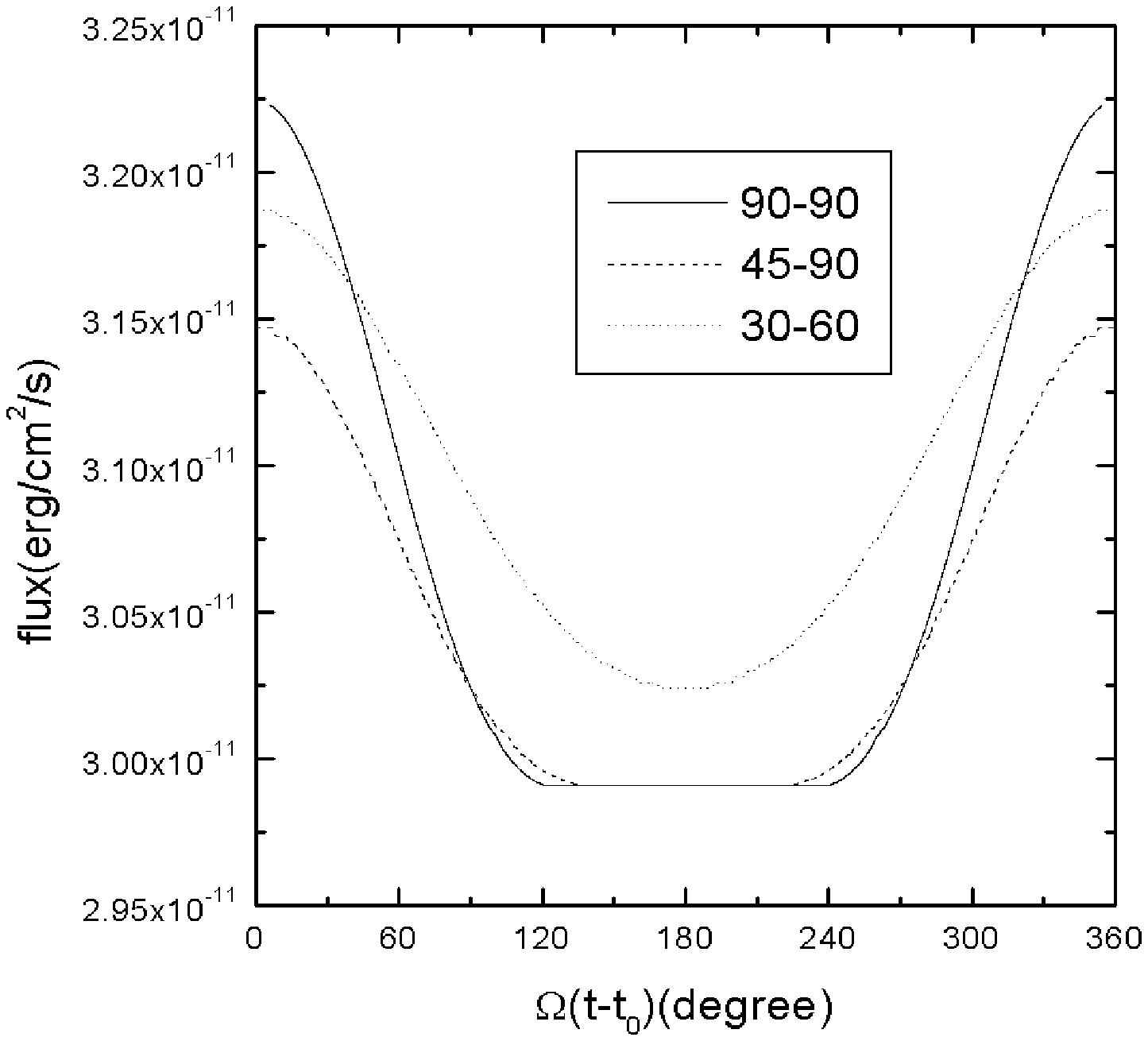}
\noindent{Fig.~\ref{fig:UTC8E42R13light418}}
\end{figure}

\begin{figure}[ht]
\vbox to7.2in{\rule{0pt}{7.2in}}
\includegraphics{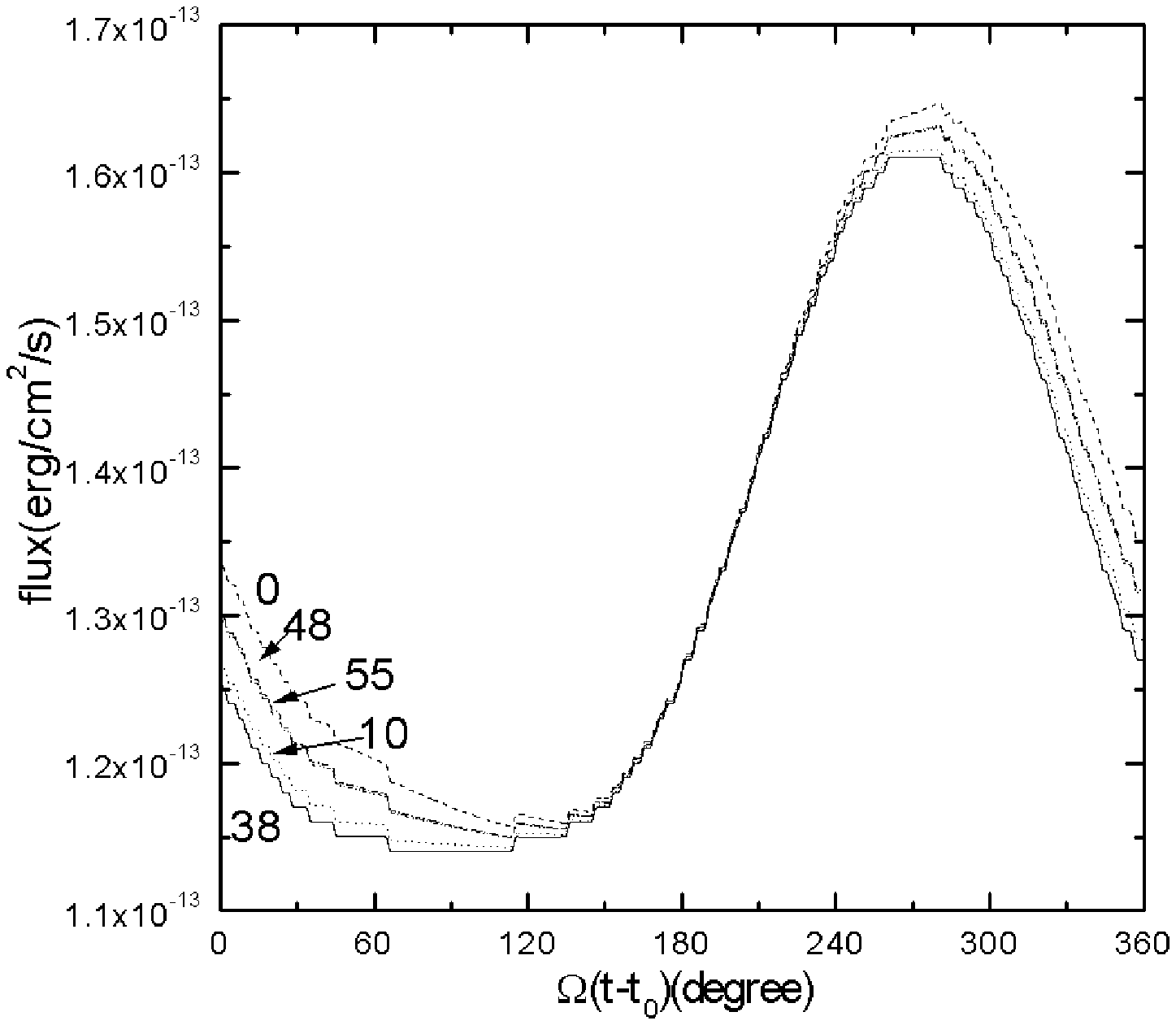}
\noindent{Fig.~\ref{fig:spotC7B312E40R12}}
\end{figure}

\begin{figure}[ht]
\vbox to7.2in{\rule{0pt}{7.2in}}
\includegraphics{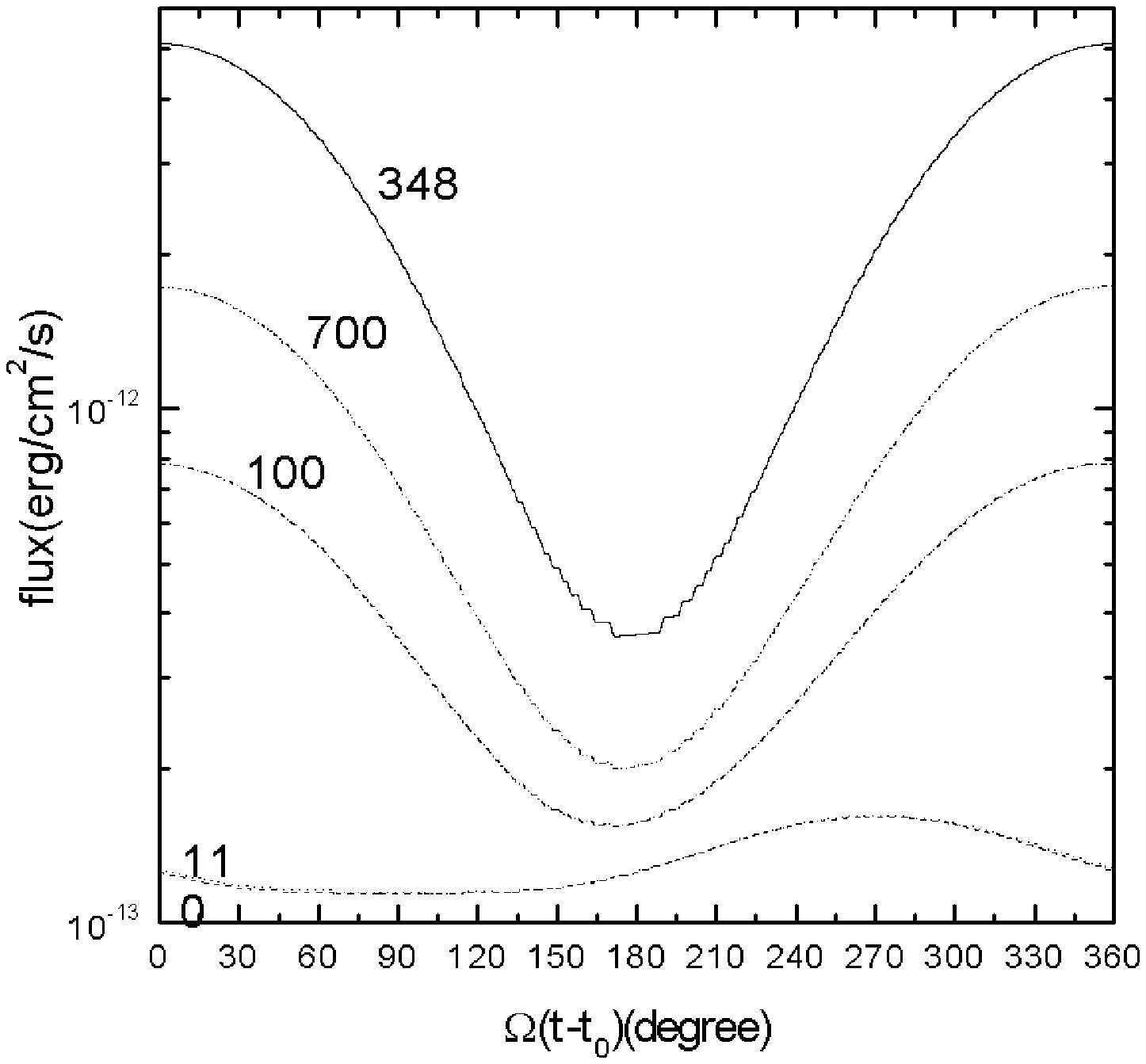}
\noindent{Fig.~\ref{fig:spotC7B312E42R12}}
\end{figure}

\begin{figure}[ht]
\vbox to7.2in{\rule{0pt}{7.2in}}
\includegraphics{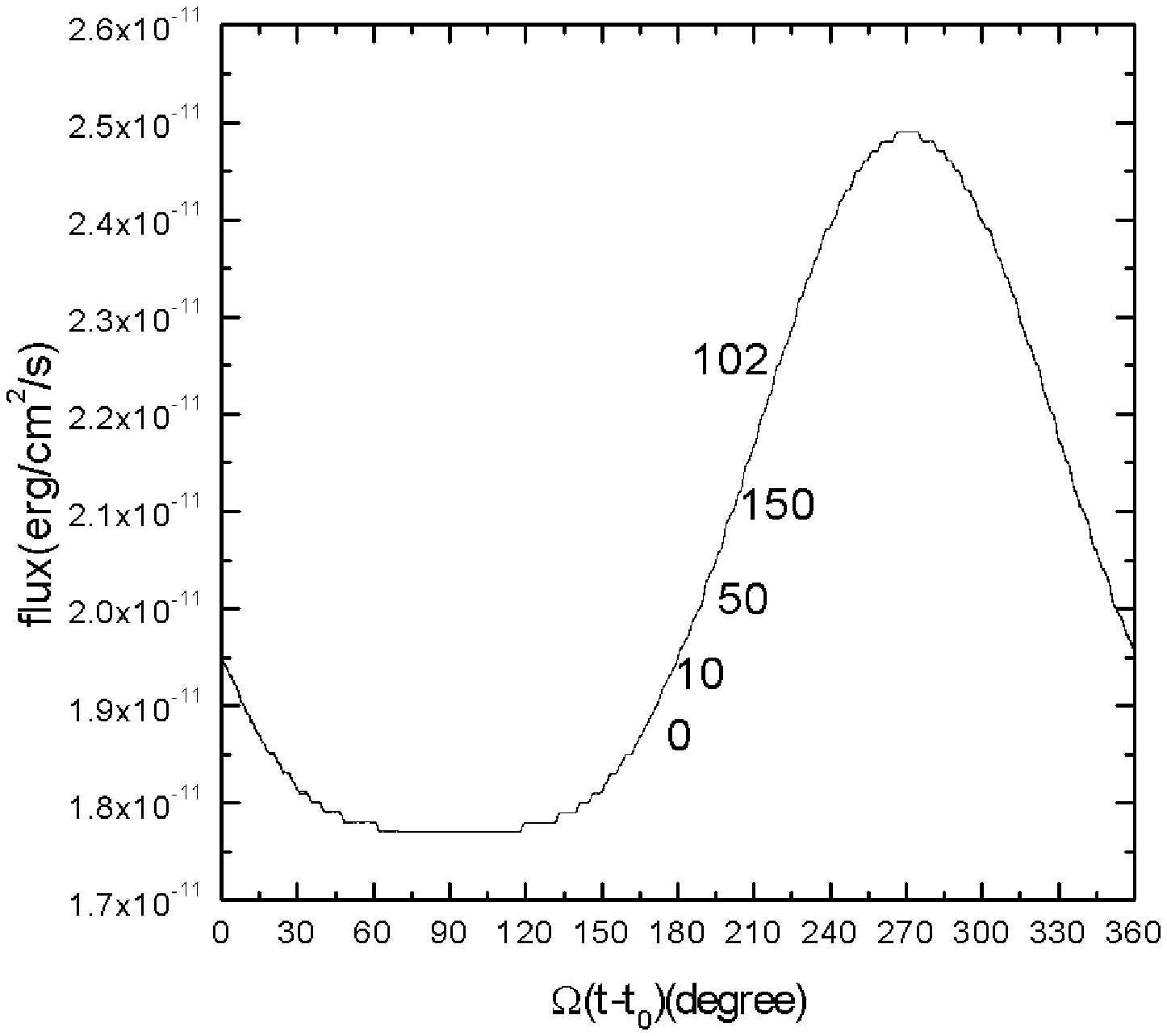}
\noindent{Fig.~\ref{fig:spotC8B312E40R12}}
\end{figure}

\begin{figure}[ht]
\vbox to7.2in{\rule{0pt}{7.2in}}
\includegraphics{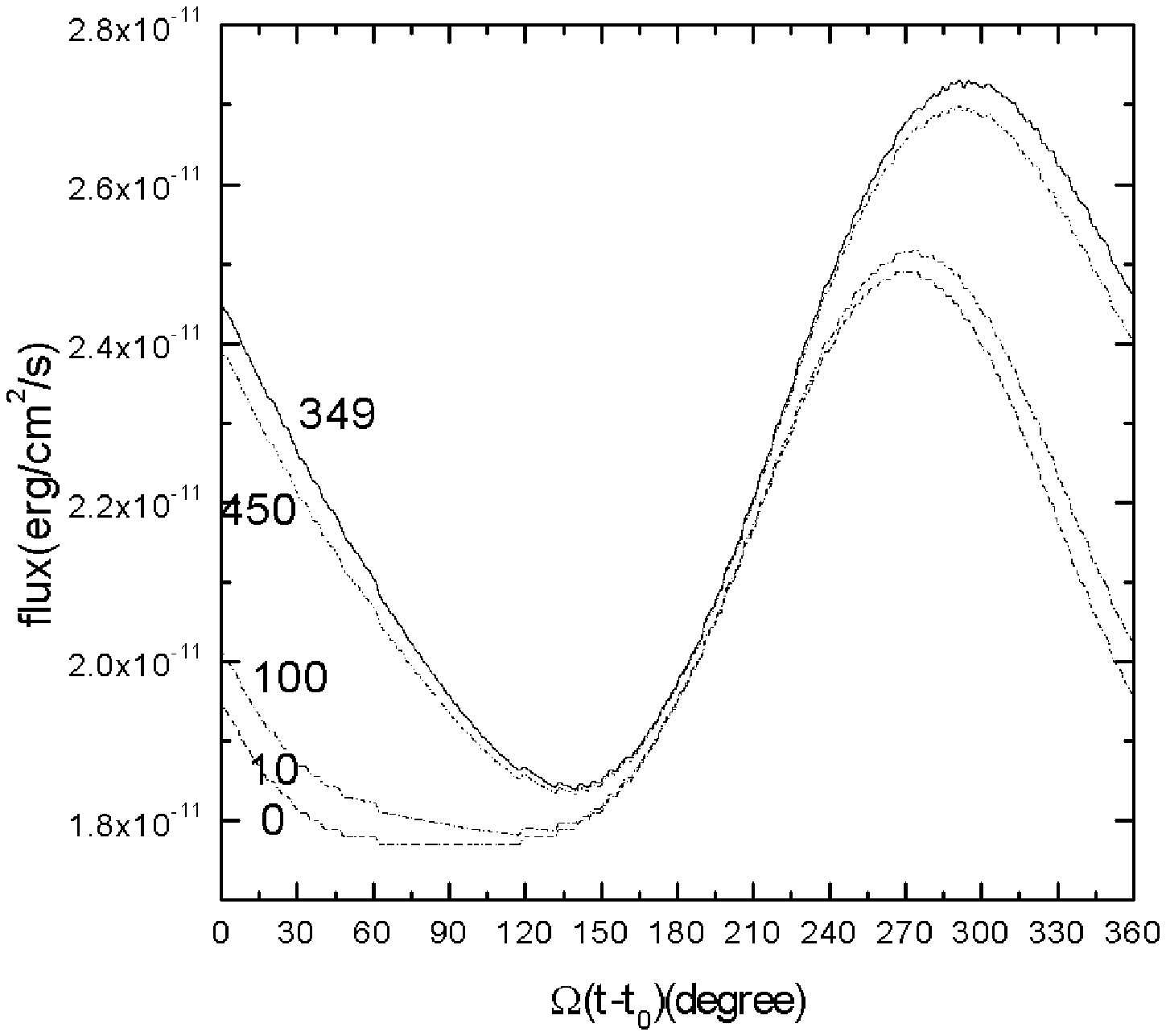}
\noindent{Fig.~\ref{fig:spotC8B312E42R12}}
\end{figure}

\begin{figure}[ht]
\vbox to7.2in{\rule{0pt}{7.2in}}
\includegraphics{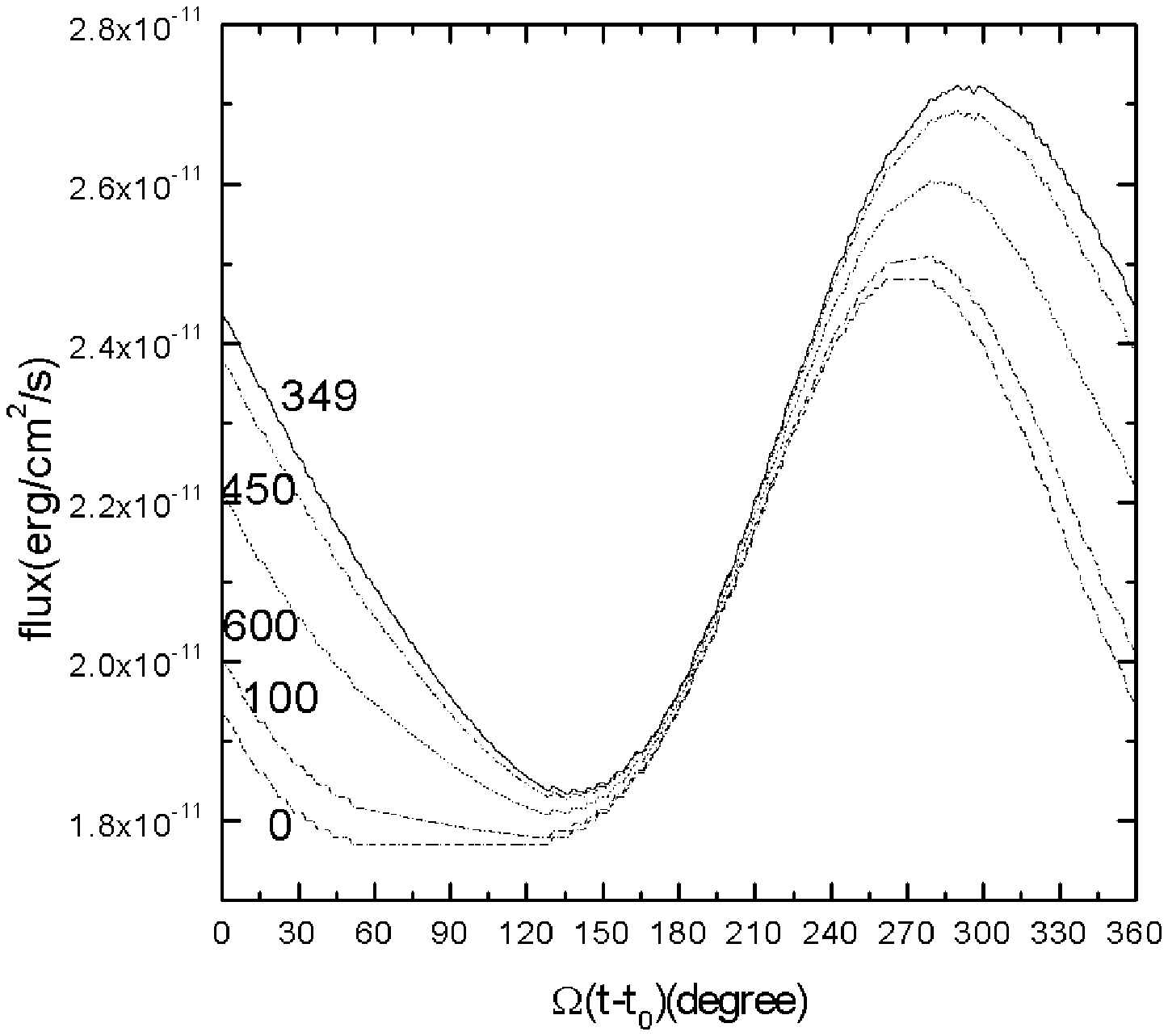}
\noindent{Fig.~\ref{fig:spotC8B15E42R12}}
\end{figure}

\begin{figure}[ht]
\vbox to7.2in{\rule{0pt}{7.2in}}
\includegraphics{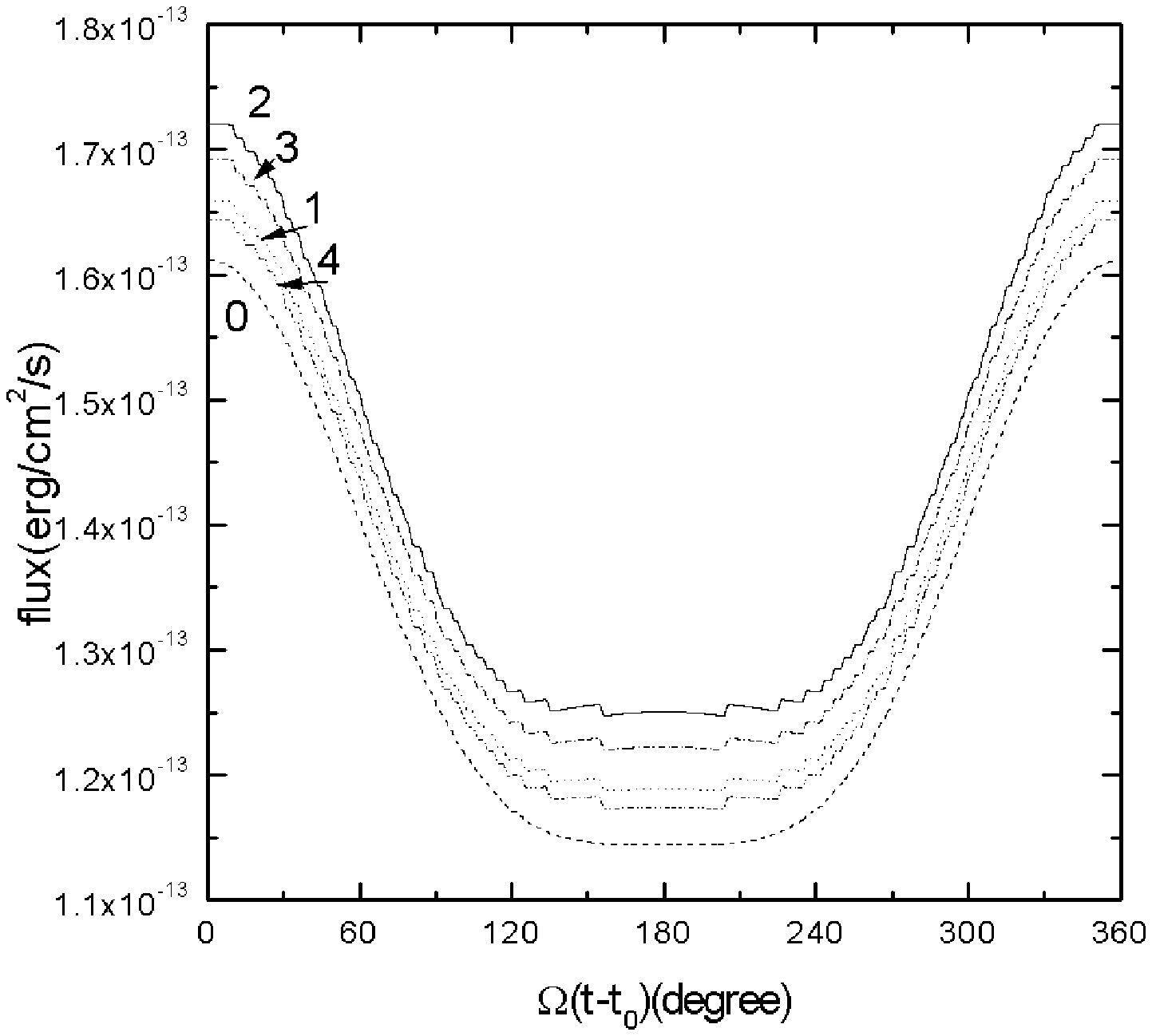}
\noindent{Fig.~\ref{fig:ringC7B312E40R12}}
\end{figure}

\begin{figure}[ht]
\vbox to7.2in{\rule{0pt}{7.2in}}
\includegraphics{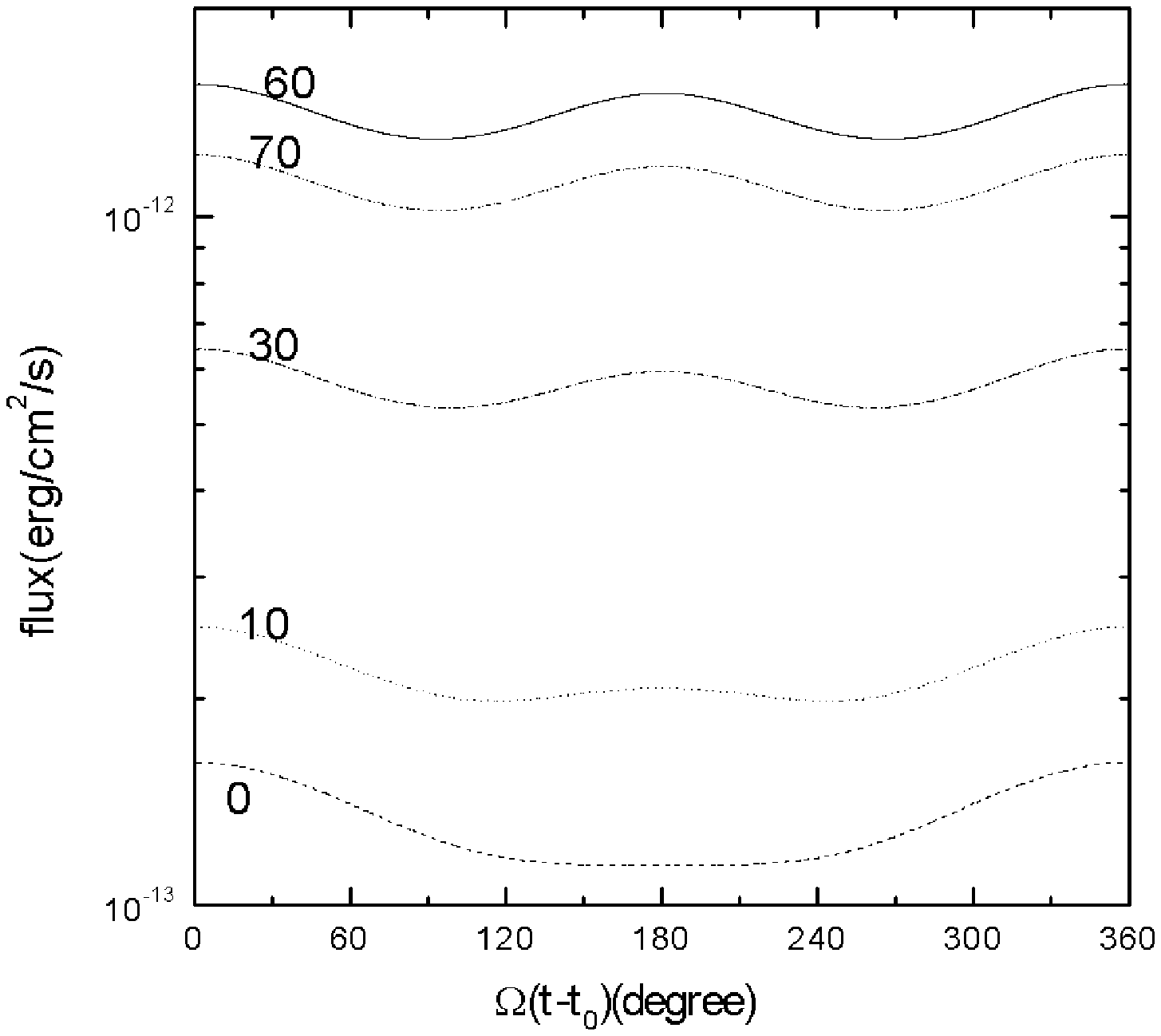}
\noindent{Fig.~\ref{fig:ringC7B312E42R12}}
\end{figure}

\begin{figure}[ht]
\vbox to7.2in{\rule{0pt}{7.2in}}
\includegraphics{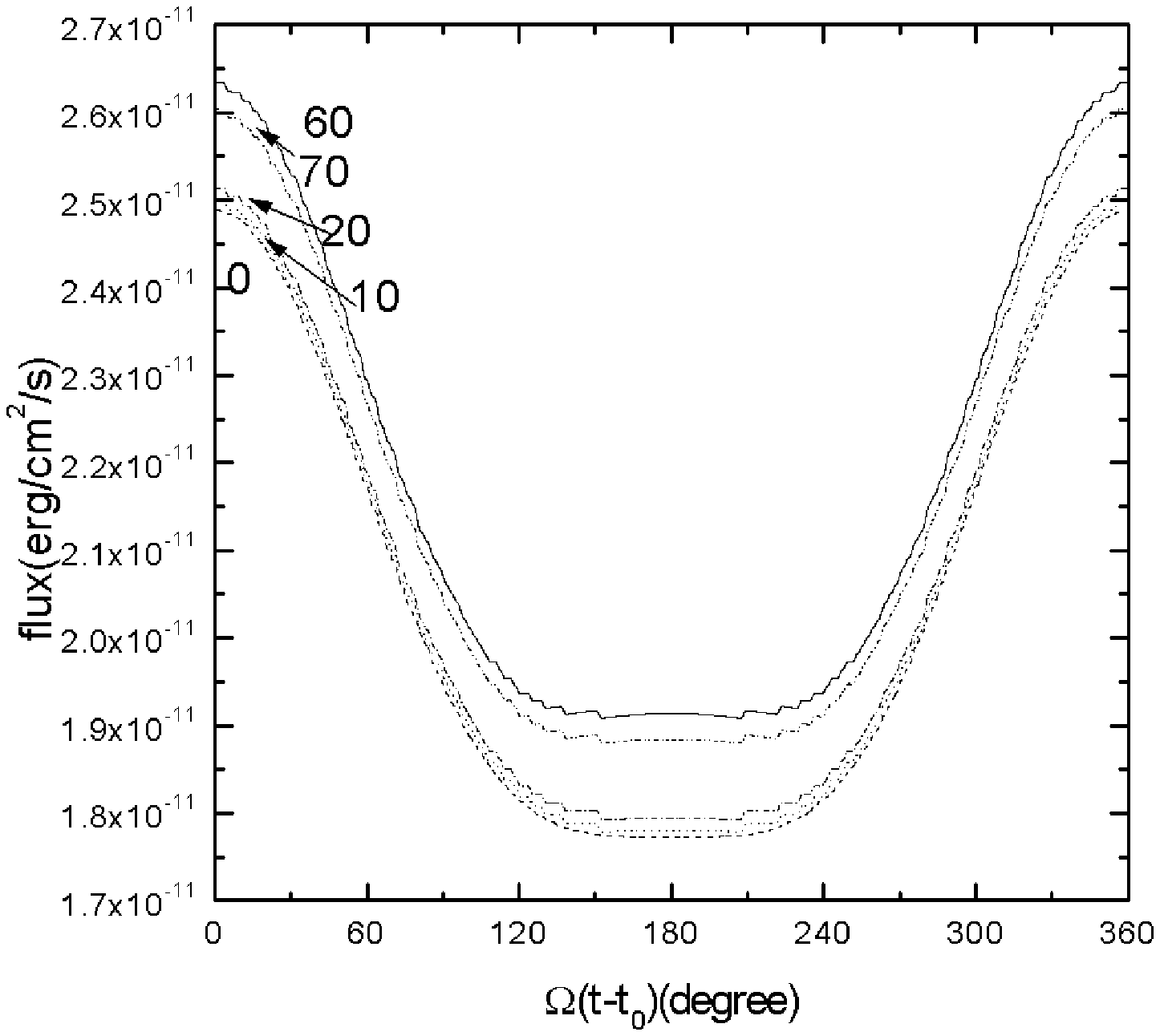}
\noindent{Fig.~\ref{fig:ringC8B312E42R12}}
\end{figure}
\end{document}